\documentclass[preprint]{aastex}
\usepackage{natbib}
\slugcomment{draft version  \today}
\usepackage[section] {placeins}
\usepackage{longtable}
\usepackage{lscape}

\begin{document}

\title{Planet Hunters: A Transiting Circumbinary Planet in a Quadruple Star System}
\author{Megan E. Schwamb\altaffilmark{1,}\altaffilmark{2,17}, Jerome A. Orosz\altaffilmark{3}, Joshua A. Carter\altaffilmark{4,18},
 William F. Welsh\altaffilmark{3}, Debra A. Fischer\altaffilmark{5}, Guillermo Torres\altaffilmark{4}, Andrew W. Howard\altaffilmark{6}, Justin R. Crepp\altaffilmark{7,8}, William C. Keel\altaffilmark{9,10}, Chris J. Lintott\altaffilmark{11,12}, Nathan A. Kaib\altaffilmark{13,19}, Dirk Terrell\altaffilmark{14}, Robert Gagliano\altaffilmark{15}, Kian J. Jek\altaffilmark{15}, Michael Parrish\altaffilmark{12}, Arfon M. Smith\altaffilmark{12}, Stuart Lynn\altaffilmark{12}, Robert J. Simpson\altaffilmark{11}, Matthew J. Giguere\altaffilmark{5}, and Kevin Schawinski\altaffilmark{1,}\altaffilmark{2,16,20} }

\altaffiltext{1}{Yale Center for Astronomy and Astrophysics, Yale University,P.O. Box 208121, New Haven, CT 06520, USA}
\altaffiltext{2}{Department of Physics, Yale University, New Haven, CT 06511, USA}
\altaffiltext{3}{Department of Astronomy, San Diego State University, 5500 Campanile Drive, San Diego, CA 92182-1221, USA}
\altaffiltext{4}{Harvard-Smithsonian Center for Astrophysics, 60 Garden Street, Cambridge, MA 02138, USA}
\altaffiltext{5}{Department of Astronomy, Yale University, New Haven, CT 06511, USA}
\altaffiltext{6}{Institute for Astronomy, University of Hawaii, 2680 Woodlawn Drive, Honolulu, HI 96822, USA}
\altaffiltext{7}{Department of Astronomy, California Institute of Technology, 1200 E.California Blvd.,Pasadena, CA91125, USA}
\altaffiltext{8}{Department of Physics, University of Notre Dame, 225 Nieuwland Science Hall, Notre Dame, IN 46556, USA}
\altaffiltext{9}{Department of Physics $\&$ Astronomy, 206 Gallalee Hall, 514 University Blvd., University of Alabama, Tuscaloosa, AL 35487-034, USA}
\altaffiltext{10}{SARA Observatory, Dept. of Physics and Space Sciences, Florida Institute of Technology, Melbourne, FL 322901, USA}
\altaffiltext{11}{Oxford Astrophysics, Denys Wilkinson Building, Keble Road, Oxford OX1 3RH, UK}
\altaffiltext{12}{Adler Planetarium, 1300 S. Lake Shore Drive, Chicago, IL 60605, USA}
\altaffiltext{13}{Northwestern University, 2131 Tech Drive, Evanston, IL, 60208, USA}
\altaffiltext{14}{Department of Space Studies, Southwest Research Institute, Boulder, Colorado 80302, USA}
\altaffiltext{15} {Planet Hunters}
\altaffiltext{16}{Institute for Astronomy, Department of Physics, ETH Zurich, Wolfgang-Pauli-Strasse 16, CH-8093 Zurich, Switzerland}
\altaffiltext{17}{NSF Astronomy and Astrophysics Postdoctoral Fellow}
\altaffiltext{18}{Hubble Fellow}
\altaffiltext{19} {CIERA Postdoctoral Fellow}
\altaffiltext{20}{Einstein Fellow}
\email{megan.schwamb@yale.edu}

\begin{abstract}

We report the discovery and confirmation of a transiting circumbinary planet (PH1b) around KIC 4862625, an eclipsing binary in the \emph{Kepler} field. The planet was discovered by volunteers searching the first six Quarters of publicly available \emph{Kepler} data as part of the Planet Hunters citizen science project. Transits of the planet across the larger and brighter of the eclipsing stars are detectable by visual inspection every $\sim$137 days, with seven transits identified in Quarters 1-11. The physical and orbital parameters of both the host stars and planet were obtained via a photometric-dynamical model, simultaneously fitting both the measured radial velocities and the \emph{Kepler} light curve of KIC 4862625. The 6.18 $\pm$ 0.17 R$_{\oplus}$ planet orbits outside the 20-day orbit of an eclipsing binary consisting of an F dwarf ( 1.734 $\pm$  0.044 R$_\odot$,  1.528 $\pm$ 0.087 M$_\odot$) and M dwarf ( 0.378 $\pm$ 0.023 R$_\odot$, 0.408 $\pm$ 0.024 M$_\odot$). For the planet, we find an upper mass limit of 169 M$_{\oplus}$ (0.531 Jupiter masses) at the 99.7$\%$ confidence level. With a radius and mass less than that of Jupiter, PH1b is well within the planetary regime. Outside the planet's orbit, at $\sim$1000 AU, a previously unknown visual binary has been identified that is likely bound to the planetary system, making this the first known case of a quadruple star system with a transiting planet.

\end{abstract}

\section{Introduction} 

For the past three and a half years NASA's \emph{Kepler} spacecraft \citep{2010Sci...327..977B}  has monitored, nearly continuously, over 160,000 stars looking for the signatures of transiting exoplanets. The \emph{Kepler} team has discovered more than 2300 planet candidates \citep{2012arXiv1202.5852B}, nearly quadrupling the sample of known planets, in addition to identifying close to 2200 eclipsing binary stars \citep{2011AJ....141...83P,2011AJ....142..160S}. One of \emph{Kepler's} many successes has been the first confirmed identification of transiting circumbinary planets, planets orbiting both stars in a binary star system \citep{2011Sci...333.1602D, 2012Natur.481..475W,2012Sci...337.1511O,2012ApJ...758...87O}. Although planets have previously been suspected to be orbiting both stars in a binary star system before \emph{Kepler} 
\citep{1999Natur.402...57B,2008A&A...480..563D,2009AJ....137.3181L,2012A&A...543A.138B}, the \emph{Kepler} era ushered in the first confirmed identification of such systems.

Identifying transiting exoplanets in binary star systems, on P-type (circumbinary) orbits where the planet orbits around both stars, is difficult when using the automated transit search methods and candidate validation processes employed by the \emph{Kepler} team. The Transiting Planet Search (TPS) \citep{2012arXiv1201.1048T,2010ApJ...713L..87J, 2002ApJ...564..495J} pipeline and the candidate vetting process were primarily designed to identify transiting planets around single stars \citep{2012arXiv1202.5852B}. Transits due to planets in binary star systems can easily be drowned out by the much larger stellar eclipse signal and may even be superimposed on top of a primary or secondary eclipse escaping detection. What  makes automatic detection challenging is the fact that the planet transits in such systems are not necessarily constant in duration or repeated at regular intervals due to the changing positions and velocities of the stars at each transit. Alternative detection techniques, including visual inspection and the search for eclipse timing variations on the host stars due to non-Keplerian three-body effects, are currently being employed to find planets in the nearly 750 eclipsing binary systems with orbital periods larger than 1 day. These efforts have proven successful with the discovery of six transiting circumbinary planets in the \emph{Kepler} field \citep[Kepler-16b, Kepler-34b, Kepler-35b, Kepler-38b, Kepler-47a, Kepler-47b;][]{2011Sci...333.1602D, 2012Natur.481..475W,2012Sci...337.1511O,2012ApJ...758...87O}. 

These multistar planetary systems are the extremes of planet formation and serve as the test grounds for planetary accretion and evolution models, that must be able to explain the existence of planets in these dynamically challenging environments. \cite{2012ApJ...754L..16P} and \cite{2012ApJ...752...71M} are unable to simulate the creation of planetary cores at the distances the \emph{Kepler} circumbinary planets currently reside suggesting these planets formed further from their parent stars and migrated inward to their present orbits. However, \cite{2012ApJ...761L...7M} finds in the outer regions of the disk between 4-10 AU, where we would then expect these planets to have formed, planetesimal growth is also inhibited due to turbulence in the planetesimal and gas disk. Models of circumbinary protoplanetary disk evolution by \cite{2012ApJ...757L..29A} predict that disks around close binaries with semimajor axes ($a$) less than 1 AU live longer than those around single stars, but disk lifetimes decline as photoevaporation increases at larger binary semimajor axes. As a result, they predict a dearth of circumbinary planets around wide binaries with $a > 10$ AU and an abundance of circumbinary planets in stellar binaries with $a <$ 1 AU. \cite{2000MNRAS.317..773B} predict that circumstellar disks around binaries with $a <$ 100 AU should be coplanar to the orbit of the binary. Thus any planets formed from such a disk are expected to all be aligned with the binary's orbit. According to dynamical evolution models by \cite{2008A&A...483..633P}, Jupiter-mass planets on circumbinary orbits will be rare and difficult to find. In their simulations, the Jupiter-mass bodies typically had a close encounter with the secondary star during migration causing the planet to be subsequently ejected from the system or scattered outward to the edge of the circumstellar disk. Therefore planets with masses equal to or less than that of Saturn are predicted to be the most abundant in these types of planetary systems. By finding more planets in multistar systems we can probe the environments these planets reside in and begin to test and evaluate the predictions from these formation and dynamical evolution models. 

We report here the discovery and characterization of a new transiting circumbinary planet, orbiting a known \emph{Kepler} eclipsing binary, via visual inspection of the first 16 months of publicly released \emph{Kepler} data by volunteers from the Planet Hunters citizen science project. This is Planet Hunters' first confirmed planet, and we subsequently refer to the planet as `PH1b'. The 6.18 $\pm$ 0.17 R$_{\oplus}$ planet orbits outside a 20-day period eclipsing binary. Beyond the orbit of PH1b, a distant visual binary located at a  projected distance of $\sim$1000 AU is likely bound to the planetary system. This is the first identification of a confirmed transiting planet orbiting in a quadruple star system. In this Paper, we present the discovery, observational follow-up, and characterization of this hierarchical stellar-planetary system. Combining the \emph{Kepler} photometric light curve and radial velocity observations with a 3-body photometric-dynamical model, we constrain the physical and orbital properties of the planet and host stars.

\section{Discovery}

The Planet Hunters\footnote{http://www.planethunters.org} \citep{2012MNRAS.419.2900F,2012ApJ...754..129S}  citizen science project uses the power of human pattern recognition via the World Wide Web to identify transits in the \emph{Kepler} public data. Planet Hunters uses the Zooniverse\footnote{http://www.zooniverse.org} platform, first described in \citep{2008MNRAS.389.1179L,2011MNRAS.412.1309S}, to present visitors to the Planet Hunters website with a randomly selected $\sim$30-day light curve segment from one of \emph{Kepler's} $\sim$160,000 target stars. 5-10 independent volunteers review each 30-day light curve segment and draw boxes directly on the web interface to mark the locations of visible transits. For further details of the website and classification interface we refer the reader to \citep{2012ApJ...754..129S}. 

In addition to the main classification interface, the Planet Hunters website hosts an object-oriented discussion and investigation tool known as `Planet Hunters Talk' (referred to as `Talk')\footnote{http://talk.planethunters.org - The code is available under an open-source license at https://github.com/zooniverse/Talk}. Talk hosts forum-style message boards enabling volunteers to discuss and analyze interesting light curves together. In addition each 30-day light curve presented on the Planet Hunters main classification interface has a dedicated page on the Talk website where volunteers can write comments, add searchable Twitter-like hash tags, and group similar light curves together. Volunteers are directed to these pages through the main Planet Hunters website. Once a classification on the main interface is complete, the volunteer is asked ``Would you like to discuss this star?". If the volunteer answers `yes', he or she is directed to the light curve's dedicated Talk page. From there a Planet Hunters volunteer can also view all available public light curve data for that \emph{Kepler} target presented as it would be seen on the main Planet Hunters website, including the ability to zoom in both time and brightness. 

Talk was designed to aid in identifying objects of particular interest and unusual light curves that are difficult to find via automatic detection algorithms and beyond the scope of the main Planet Hunters classification interface. It also provides an avenue for communicating directly with the Planet Hunters science team through the message boards and comments. From 2010 December 16 to 2012 June 9, in total, 69 percent of registered Planet Hunters participants have visited the Talk site with 23 percent having contributed comments. In this time frame, an average of 116 and a median of 72 Planet Hunters users visit Talk each day. In addition an average of 511 and median of 329 comments discussing presented light curves are generated on Talk daily. We note that the two planet candidates presented in \cite{2012arXiv1202.6007L} were first identified by the Planet Hunters science team via the discussion of the light curves on the Talk message boards before a systematic search of the Q2 light curves had been completed.

\begin{figure}
\centering
\epsscale{1.08}
\plotone{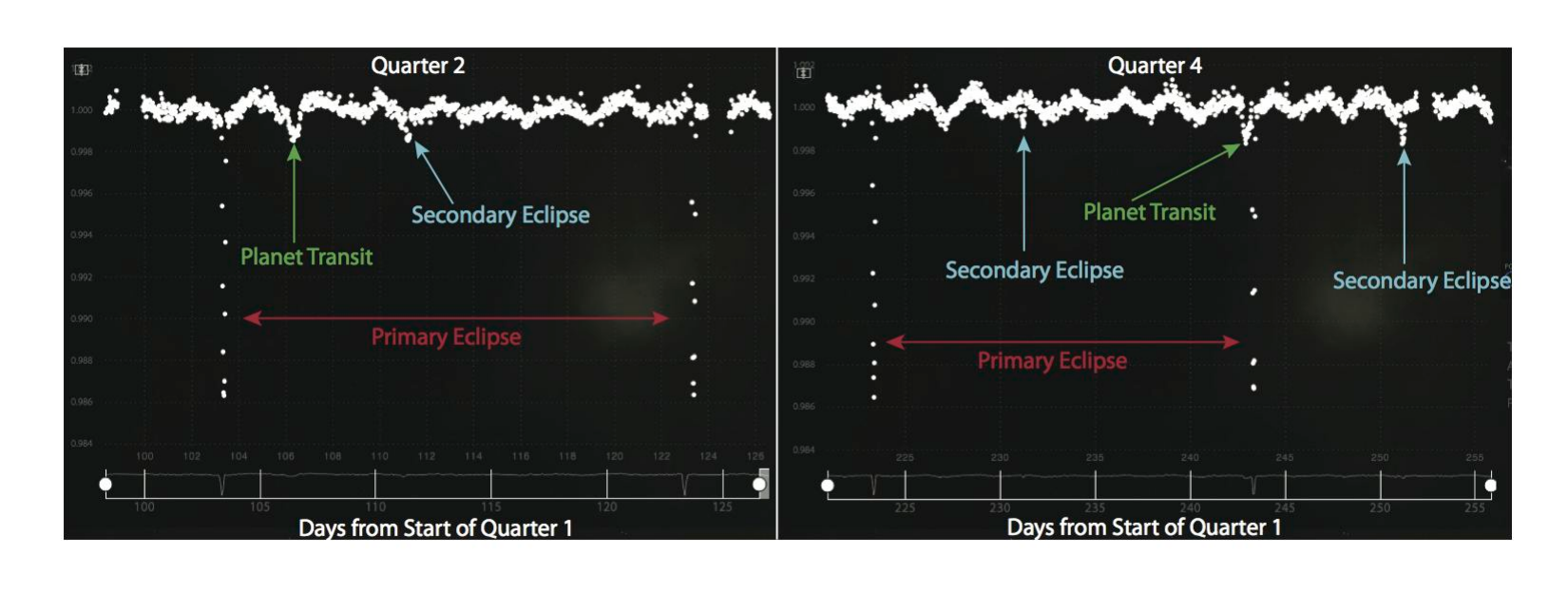}
\caption{Planetary transits  in addition to binary-star eclipses identified by Planet Hunters volunteers in KIC 4862625 as viewed in the Planet Hunters interface.}
\label{fig:PHinterace}
\end{figure}

On 2011 May 11, coauthor Kian Jek posted on Talk highlighting the light curve of eclipsing binary KIC 12644769, identifying additional transits due to a possible third body. \cite{2011AJ....142..160S} had already noted the additional transit features, and the system subsequently turned out to be Kepler-16, the first circumbinary discovered in the \emph{Kepler} field \citep{2011Sci...333.1602D}. After the announced discovery of Kepler-16b, the Planet Hunters science team posted links to the Talk pages for the entire  \cite{2011AJ....142..160S} eclipsing binary list encouraging volunteers to examine these eclipsing binary light curves and perform their own search for additional transits due to an orbiting planet. During visual inspection of the light curve for detached eclipsing binary KIC 4862625, coauthor Robert Gagliano spotted two transit-like features in Quarters 2 and 4 (see Figure \ref{fig:PHinterace}) separated by $\sim$137 days, and on 2012 March 2 highlighted this light curve on the Talk forums as a potential circumbinary. At the time only Quarters 1-4 ($\sim$310 days of science data) were available on Planet Hunters website, but Quarters 1-6 were publicly available. On 2012 March 3, Kian Jek, examined the Quarter 5 and 6 light curves, obtained from the \emph{Kepler} Mission Archive\footnote{http://archive.stsci.edu/kepler} hosted by the  Mikulski Archive for Space Telescopes (MAST\footnote{http://archive.stsci.edu}), identifying a third transit with similar depth in Quarter 5 occurring roughly 137 days after the Quarter 4 transit. These additional transit events in Quarters 1-5 as well as others subsequent Quarters were also independently identified by \cite{2012arXiv1210.3850K} in a separate and concurrent analysis of the \emph{Kepler} public data. After examining the reported column pixel centroids in the FITS files for a shift during in and out of transit that may indicate contamination from a faint background eclipsing binary, Kian Jek noted the presence of this third transit in a Talk discussion thread\footnote{http://talk.planethunters.org/discussions$/$DPH1014m5m?group$\_$id=SPH10052872} and notified the Planet Hunters science team of the potential discovery of a circumbinary planet.

\section{\emph{Kepler} Light Curve}
\label{sec:lightcurve}
There are two types of \emph{Kepler} light curves provided in MAST: Pre-Search Data Conditioning (PDC) and Simple Aperture Photometry (SAP). The PDC correction optimizes the light curve for detection for planet transits as well as removing systematic effects and long-term trends, but also makes some assumption for stellar variability and attempts to remove that signal. The SAP light curve is the summed  flux in the target aperture with no further corrections. For further specifics of the \emph{Kepler} data processing and data products we refer the reader to \cite{2010ApJ...713L..87J,2010ApJ...713L.120J,Keplerdatamanual}, and \cite{2012PASP..124.1000S}. We note that the original discovery on the Planet Hunters website was using the PDC-corrected light curve, but for the analysis presented here, we choose to use the SAP light curves in order to preserve the stellar variability and the eclipse signals. 

In the initial stages of trying to determine whether transits were occurring, we took a simple approach to reduce the  light curve to remove brightness changes due to  stellar variability. Individual sets of raw photometry with potential transits in Quarters 2,4, and 5 were fitted with a combination of linear and periodic (sine) terms. The fits were done in multiple steps, first using a linear term, then linear plus periodic terms until the fit was no longer improving. Since the stellar variability had a much higher frequency, this approach worked well to remove the linear and low frequency terms, leaving the transit (and any stellar eclipses) in each data set clearly visible. The multiple transits were then compared visually with one another, showing similar depth ($\sim$0.1$\%$) consistent with the third body transiting being planetary in size, $\sim$2.8 R$_{\oplus}$ assuming a primary (Aa) stellar radius of 0.804 R$_{\odot}$ as reported in the {Kepler} Input Catalog\footnote{http://archive.stsci.edu/Kepler/Kepler\underline{\hspace*{0.25cm}}fov/search.php}  (KIC, \citeauthor{2011AJ....142..112B} \citeyear{2011AJ....142..112B}).

\begin{figure}
\epsscale{1.0}
\plotone{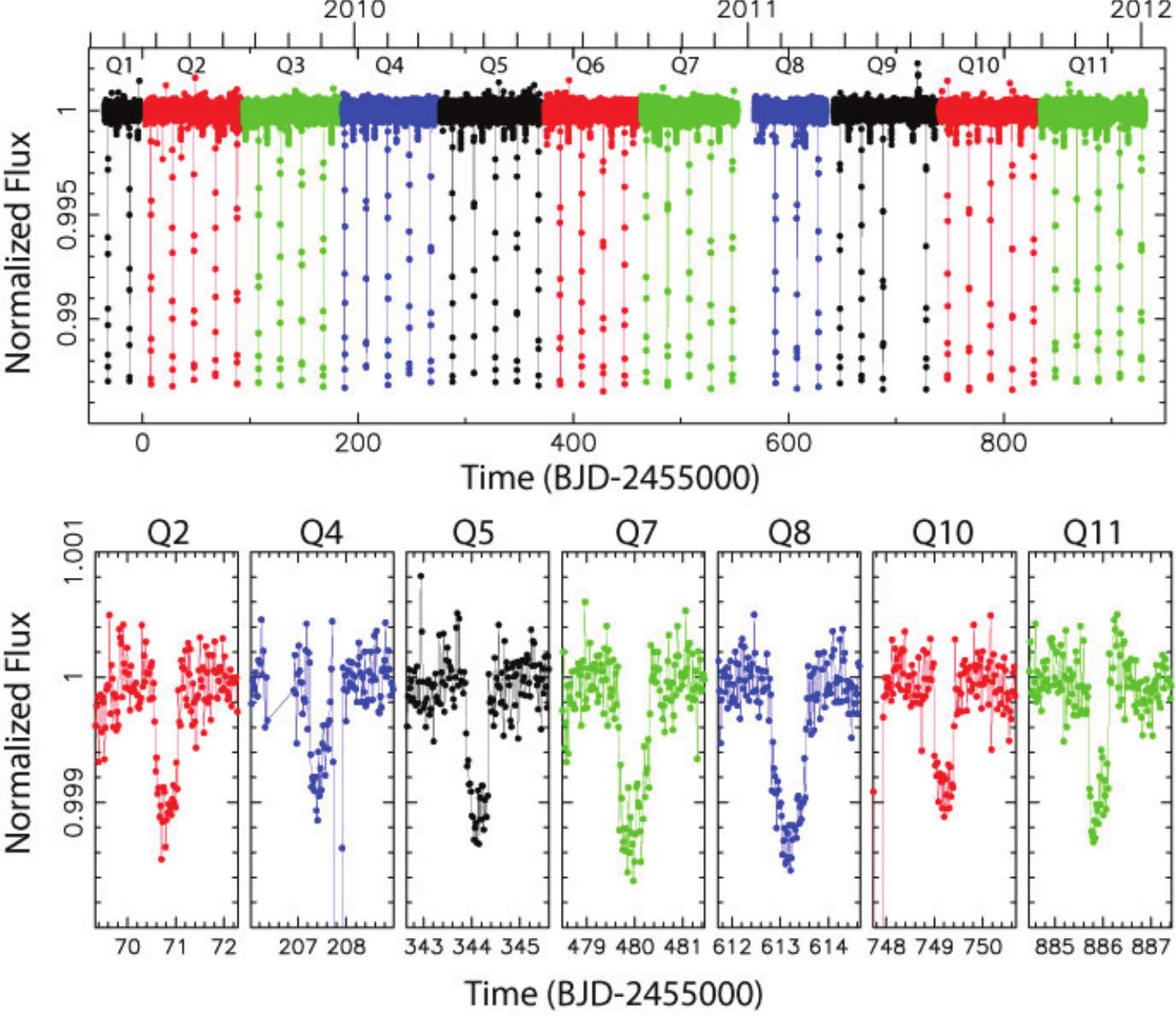}
\caption{ Reduced Kepler light curve for KIC 4862625 for Quarters 1-11 used to measure stellar eclipse timing variations, timing offsets of the planet transits, and changes in planet transit duration described in Section \ref{sec:lightcurve}. Top: Full light curve for  KIC 4862625 including stellar eclipse and planet transits. The colors plotted (on the online version) are used to separate the different Kepler Quarters. Bottom: Isolated individual identified transits of the planet across star Aa. In Quarter 4 (Q4), the planet transit occurs just before a primary eclipse. }
\label{fig:lightcurve}
\end{figure}

For the more detailed analysis presented in this Section, we perform our own detrending and removal of instrumental systematics using a method adapted from that \cite{2012ApJ...761..157B}. For each individual Quarter of \emph{Kepler} observations, the light curve is further subdivided in sections by data gaps and discontinuities. Each light curve section is normalized by a continuum calculated by masking out the stellar eclipses and planet transits and then fitting a high order cubic spline to the remaining data points using an iterative sigma-clipping routine. The resulting normalized light curve for KIC 4862625, for Quarters 1-11 is shown in Figure \ref{fig:lightcurve} and listed in Table \ref{tab:lightcurve}.  The KIC reports a magnitude of 13.72 for the eclipsing binary in the  broad \emph{Kepler} filter (Kp) \citep{2011AJ....142..112B}. We will refer to the brighter more massive star as `Aa' and use `Ab' to denote the fainter lower-mass stellar companion in the eclipsing binary. These two stars have an orbital period of 20.000214 days on an eccentric orbit with an eccentricity of 0.24181 as reported by \cite{2011AJ....142..160S}. In the $\emph{Kepler}$ light curve the depth of the primary eclipses are $\sim$1.3$\%$, and the depth of the secondary eclipses are only slightly deeper than the planet transit at $\sim$0.11$\%$. Three transits of PH1b crossing Aa are visible in Quarters 2, 4, and 5, the data publicly available at the time of discovery, with another four identified in Quarters 7, 8,10, and 11. No transits across Ab were identified. 

\begin{figure}
\epsscale{0.45}
\plotone{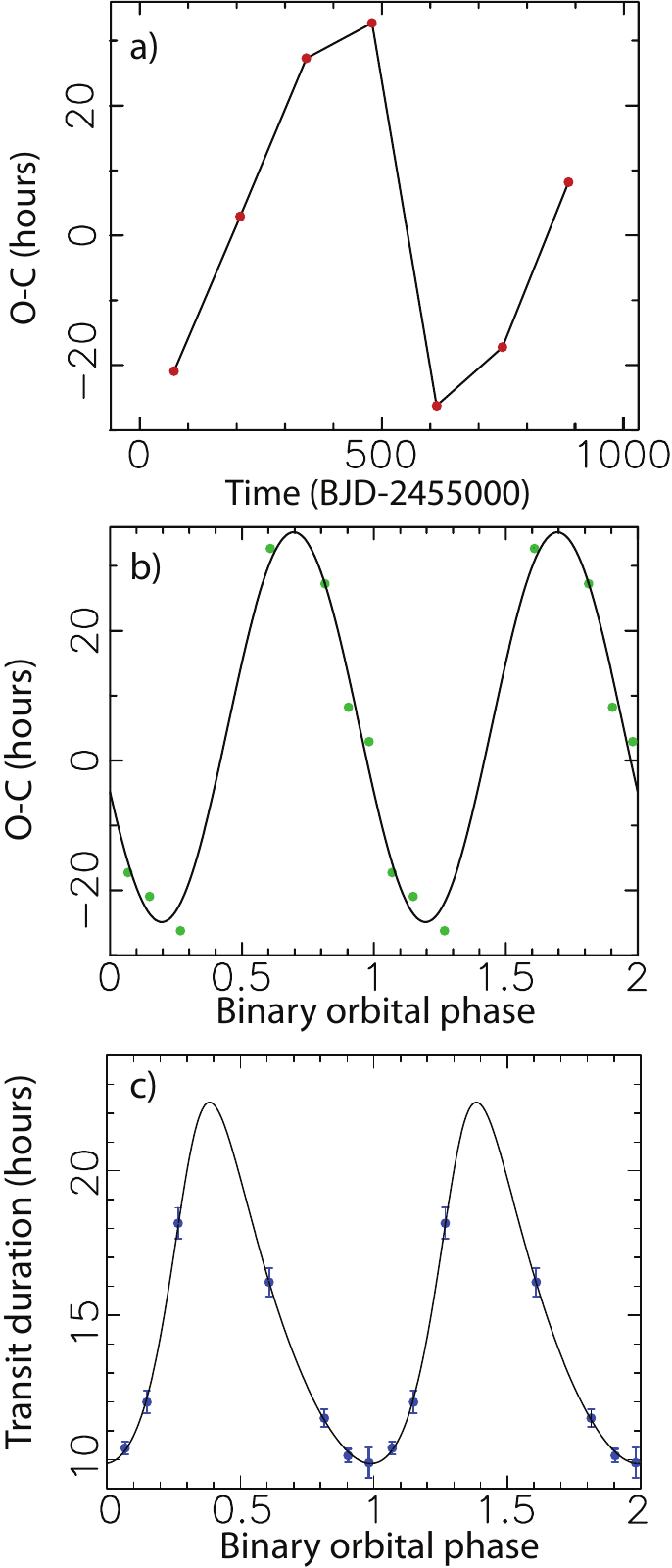}
\caption{O-C offsets and transit durations for the circumbinary planet. a) O-C offsets measured for the planet transits versus time. The 1-$\sigma$ error bars are plotted. With uncertainties on the transit times  on the order of $\sim$5 minutes, the error bars are  smaller than the size of the symbol. A solid line is used to connect the points for clarity. b) O-C offsets measured for the planet transits versus binary orbital phase. The 1-$\sigma$ error bars are plotted. With uncertainties on the transit times  on the order of $\sim$5 minutes, the error bars are  smaller than the size of the symbol. The solid line plots the best-fit sinusoid to the timing offsets. c) Transit duration as a function of binary orbital phase. Transits near secondary eclipse $\phi=0.39$ are wider than those close to primary eclipse $\phi=0.0$.  The solid line represents the durations derived from the best-fit circular Keplerian orbit.}
\label{fig:POC}
\end{figure}

One of the largest sources of false positives for \emph{Kepler} planet candidates are blended faint background eclipsing binaries, but this would require a chance sky alignment of two eclipsing binary systems along the line of sight. This scenario, though highly unlikely, cannot be completely ruled out by statistical arguments alone. If the transiting body truly is orbiting both stars in the binary, the exact timing and the duration of the planet transits should vary due to the changing positions and velocities of the stars at each transit \citep[e.g][]{1998A&A...338..479D,2000ApJ...535..338D,2011Sci...333.1602D, 2012Natur.481..475W,2012Sci...337.1511O,2012ApJ...758...87O}. A transit function \citep{2006A&A...450.1231G} is fit individually to each of the seven transits to measure the transit times and widths. The model, involving two limb-darkened spheres, is computed at an approximately one minute cadence and then binned into 29.4474 minute bins to match \emph{Kepler}'s exposure time. The measured transit midpoints are reported in Table \ref{tab:transittimes}.  The mean period is 135.648 days. We note all times and ephemerides in our analysis are using Barycentric Julian Dates  in Barycentric Dynamical Time, which we will refer to as BJD. Observed (O) minus Calculated (C) offsets and durations for the seven transits are plotted in Figure \ref{fig:POC}. O-C amplitudes as high as $\pm$ 1 day are observed and the transit durations are varying with the binary orbital phase, confirming this body is indeed orbiting stars Aa and Ab and the transit features are not due to a blended background eclipsing binary or any other astrophysical false positive. In Figure \ref{fig:POC} b, the best-fit sinusoid, reflecting a Keplerian circular orbit for the planet, to the O-C timing offsets as a function of binary phase are fit. Deviations from the expectation visible in Figure \ref{fig:POC} b, are generated by the non-zero eccentricities of both the binary and planetary orbit.

\begin{figure}
\epsscale{1.0}
\plotone{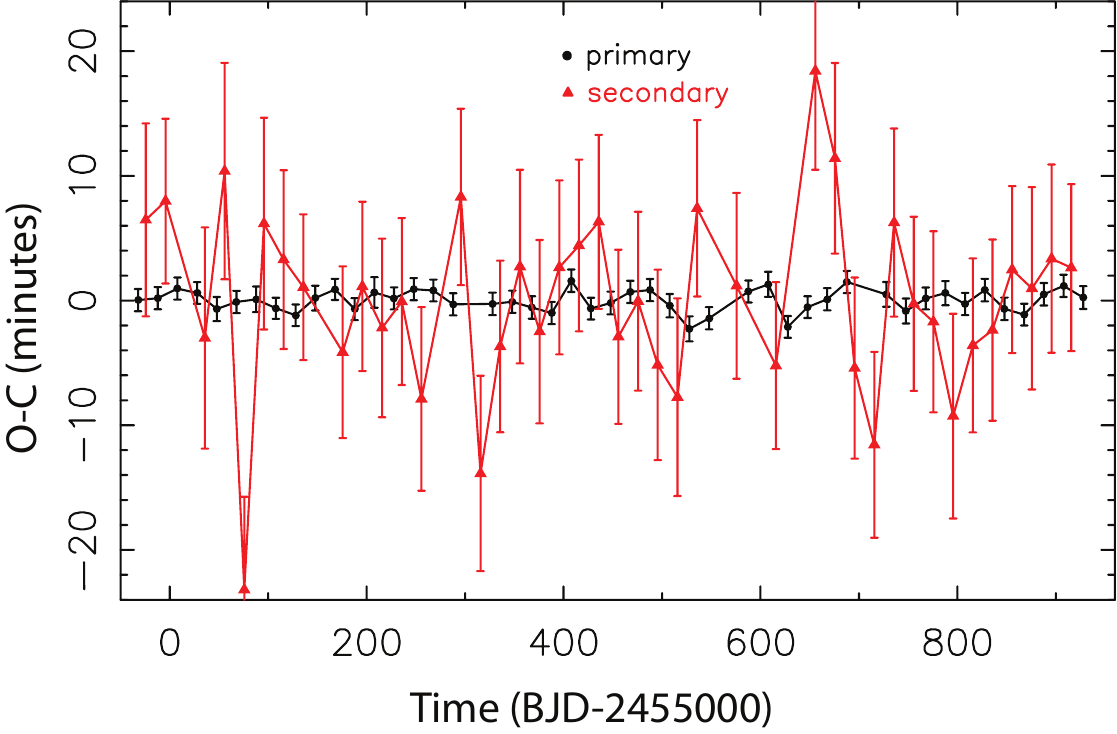}
\caption{O-C offsets measured for the primary and secondary (red in the online version)  eclipses of KIC 4862625.}
\label{fig:OC}
\end{figure}

Furthermore, the tidal field of the planet may cause nonlinear timing variations in the eclipsing binary, producing  a period difference between primary and secondary eclipses, as had been observed for the Kepler-16b, Kepler-34b, and Kepler-35b systems \citep{2011Sci...333.1602D, 2012Natur.481..475W}.  Primary and secondary eclipse timings were estimated by fitting a low-order cubic Hermite polynomial \citep[for details see][]{ 2011MNRAS.417L..31S, 2012Natur.481..475W} to each individual eclipse in the normalized Quarters 1-11 light curve. The measured primary and secondary eclipse midpoints for KIC 4862625 are reported in Table \ref{tab:eclipsetimes}. The best fitting linear primary and secondary eclipse ephemerides are: T$_0$ = -52.18091 $\pm$ 0.00018 (BJD - 2,455,000) with a period P = 20.0002498 $\pm$ 0.0000064 days for the primary eclipse and  T$_0$ = -44.3252 $\pm$ 0.0016 (BJD - 2,455,000) with P = 20.000244 $\pm$ 0.000056 days for the secondary eclipse. The difference between the primary and secondary periods is 0.50 $\pm$ 4.87 seconds. The timings of the stellar eclipses are measured and compared to the that expected from the best-fit eclipse ephemerides. The O-C offsets for the primary and secondary stellar eclipses are plotted in Figure \ref{fig:OC}. No discernible timing variation signal in either the primary or secondary eclipses were found. The planet may not be massive enough to gravitationally perturb the stellar orbits significantly over the timescale of the observations, as is the case for the Kepler-38 and Kepler-47 systems \citep{2012Sci...337.1511O,2012ApJ...758...87O}. Without detectable perturbations on the stellar orbits, we will only be able to place an upper limit on the mass of PH1b when the light curve and radial velocity observations are combined in the photometric-dynamical model fit as described in Section \ref{section:photo}. 

\section{Flux Contamination}
\label{sec:contamination}

The \emph{Kepler} pixel scale is 4$^{\prime\prime}$ per pixel, with a typical photometric aperture radius of 6$^{\prime\prime}$ \citep{2010ApJ...713L.120J}. With such a wide aperture, stellar contamination and photometric blends are a concern when analyzing the \emph{Kepler} photometry of target stars. Adding linearly, the contribution of extra light decreases the observed transit and eclipse depths. Accurately estimating the size of the transiting planet requires knowledge of the additional flux contributors to the \emph{Kepler} light curve for KIC 4862625. A search of the KIC found that two nearest sources are KIC 4862611 and KIC 4862601 located 8.68$^{\prime\prime}$ and 11.8$^{\prime\prime}$ respectively from KIC 4862625 \citep{2011AJ....142..112B}. With \emph{Kepler} magnitudes of 18.052 and 18.256 respectively, these two stars would not be completely in the \emph{Kepler} photometric aperture and at most would contribute less than a percent of total flux in the light curve.

Faint companions closer than a few arcseconds to the \emph{Kepler} target may not be found in the KIC or 2MASS catalog 
\citep{2006AJ....131.1163S,2011AJ....142..112B}. To better constrain the presence of additional stars not listed in the KIC that may be contaminating the \emph{Kepler} aperture, we obtained near-infrared adaptive optics (AO) observations using the NIRC2 imager behind the natural guide-star (NGS) AO system on the Keck II telescope \citep{2008ApJ...689.1044G}. K$^\prime$(1.948-2.299 \micron) imaging was taken on 2012 May 29 UT with NIRC2 in the narrow camera setting, yielding a 10x10$^{\prime\prime}$ field-of-view with a pixel scale of 9.9 mas per pixel. We used a 9-point dither pattern with a 2$^{\prime\prime}$ offset between each position. At each nod, 60s integrations were taken, resulting in a total integration time on target of 540s. The raw frames were dark and background subtracted, and  flat fielded. Bad and hot pixels were interpolated over. The reduced frames were then combined into a single image. Visual inspection revealed two nearby stars: a source 0.702$^{\prime\prime}$ southwest of KIC 4862625 approximately 5 times fainter than the target star and a much fainter source $\sim$3$^{\prime\prime}$ away to the southwest. From this point forward, we refer to these sources as the 0.7$^{\prime\prime}$ contaminator and the 3$^{\prime\prime}$ contaminator respectively. Both stars are well within the photometric aperture of KIC 4862625. The \emph{Kepler} data analysis pipeline would not account for the flux contribution from these contaminating stars. 

\begin{figure}
\epsscale{1.0}
\plotone{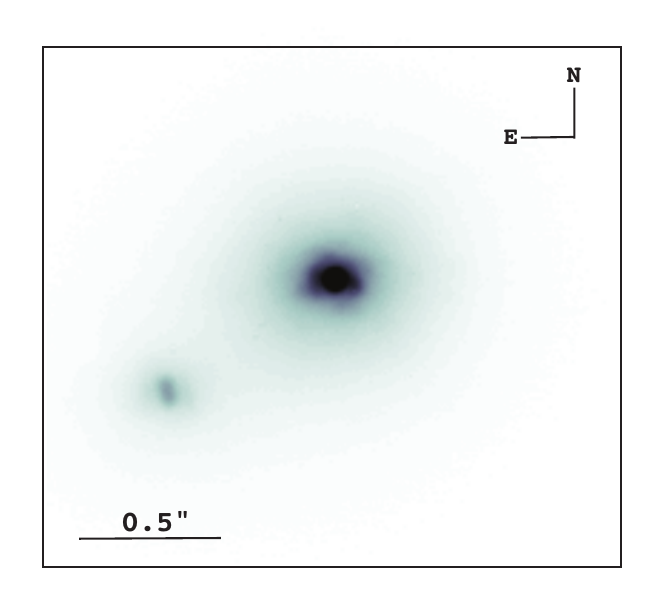}
\caption{NGS AO Ks image of KIC 4862625  and the 0.7$^{\prime\prime}$ contaminator located  0.702$^{\prime\prime}$ away from the primary Kepler target.}
\label{fig:AO}
\end{figure}

In order to estimate the light curve dilution due to the brighter 0.7$^{\prime\prime}$ contaminator, we obtained further observations with NIRC2 NGS AO in Ks (1.991-2.302 \micron) and J (1.166-1.330 \micron) filters on 2012 June 7 UT in the same observing setup. The images were reduced in the same manner as on the first night but with a 3$^{\prime\prime}$ offset between nods. J images were taken with an integration time of 90s exposures at each nod. The Ks imaging consisted of 60s at each nod. A total integration time of 810s in J and 540s in Ks respectively was achieved. There were intermittent high cirrus clouds during those observations, and therefore we were only able to obtain relative J and Ks photometry. Figure \ref{fig:AO} presents the reduced Ks AO image for the 0.7$^{\prime\prime}$ contaminator and the primary KIC target. Using aperture photometry, we measure $\delta K_s = 1.67 \pm$ 0.03 and $\delta J = 1.89 \pm 0.04 $ mag difference. 

We adopt the method developed by \cite{2012ApJ...746..123H} to extrapolate to the broad band \emph{Kepler} filter (Kp) using J and Ks colors. The measured relative J and Ks colors were combined with the reported 2MASS photometric magnitudes (J=12.714, K=12.394) to obtain the true apparent Ks and J magnitudes for both KIC 4862625 and the 0.7$^{\prime\prime}$ contaminator. The 2MASS colors are a summation of all three identified sources. The 3$^{\prime\prime}$ contaminator is too faint in AO observations to measure an accurate relative magnitude, but contributes at most at the percent level to the infrared colors. Therefore, we chose to assume KIC 4862625 and the 0.7$^{\prime\prime}$ contaminator are the only contributions to the reported 2MASS colors. The KIC reports Kp= 13.71 mag for KIC 4862625. Using the filter transforms determined by \cite{2012ApJ...746..123H} we find a Kp magnitude of 13.79 for KIC 4862625 and 16.10  for the 0.7$^{\prime\prime}$ contaminator if the primary in the Kepler target is a dwarf star. If it is a giant, we estimate a Kp magnitude of 13.77 for KIC 4862625 and 16.21  for the faint companion. Thus the 0.7$^{\prime\prime}$ contaminator contributes 11-12$\%$ of the light in the  light curve for KIC 4862625. 

It is not surprising that we find a source less than 1$^{\prime\prime}$ from KIC 4862625. \cite{2012AJ....144...42A} find that 20$\%$ of the \emph{Kepler} candidates observed with adaptive optics have a close companion within 2$^{\prime\prime}$, and \cite{2012A&A...546A..10L}report similar findings with 17$\%$ of their sample of \emph{Kepler} candidates examined had at least one contaminating source within 3$^{\prime\prime}$ of the \emph{Kepler} target. Using the KIC stellar parameters and the expected galactic extinction, KIC 4862625 is approximately 1 kpc away, thus the projected distance between the binary and the 0.7$^{\prime\prime}$ contaminator is $\sim$1000 AU. Although gravitationally bound multistellar systems exist with such separations, we would need measurements of the proper motions of both the binary (Aa and Ab) and the 0.7$^{\prime\prime}$ contaminator to confirm association. From the NGS AO observations alone, we cannot determine if the 0.7$^{\prime\prime}$ contaminator is bound to the system. The time separation of our observations is insufficient to measure proper motions, but the radial velocity observations (discussed in Section \ref{sec:rvs}), find this source has a systemic velocity consistent with that of the eclipsing binary. In addition the AO observations reveal the 0.7$^{\prime\prime}$ source is itself a binary. We further discuss the properties of this third stellar component in detail in Section \ref{sec:extrastars}.

\begin{figure}
\epsscale{1.0}
\plotone{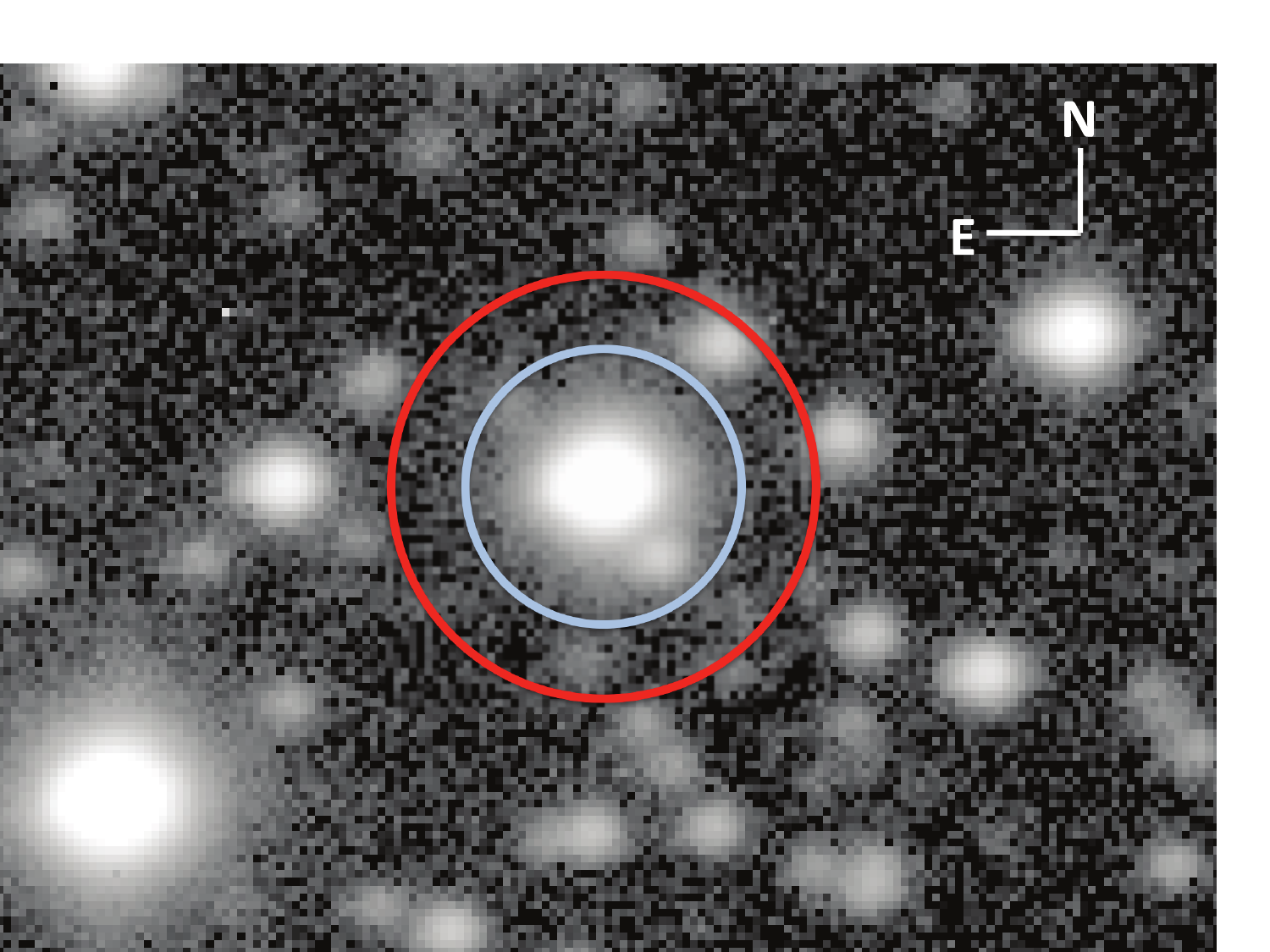}
\caption{SARA 0.9m stacked image of  KIC 4862625  and surrounding stars. The inner circle (light blue in the online version) has a radius of 6$^{\prime\prime}$ and the outer circle (red in the online version) has a radius of 10$^{\prime\prime}$ centered on KIC 4862625.}
\label{fig:sara}
\end{figure}

In order to estimate the brightness of the 3$^{\prime\prime}$ contaminator and to quantify the contributions of additional faint stars 2-6$^{\prime\prime}$ from KIC 4862625, we use R-band optical imaging from the remote SARA 1m telescope on Kitt Peak. We obtained a series of 60 60s exposures on 2012 July 18 UT, keeping individual exposures short so that all stars of interest remained in the linear count regime. The co-added SARA image is presented in Figure \ref{fig:sara}. We used both elliptical model fitting and PSF subtraction (from a similarly bright star without bright neighbors) to remove the wings of the bright-star profile. Slight irregularities in the PSF shape made the subtraction of a scaled PSF more successful. We estimate errors based on the scatter of photometry in independent fits from 6 10-exposure averages of the data. The photometric zero point was established from 20 nearby KIC stars brighter than $r$=17.5, using photometry from \cite{2011AJ....142..112B},ignoring the small color term in transforming from $R$ to \emph{Kepler} bands. The image quality was 1.5$^{\prime\prime}$ FWHM, well sampled with 0.38$^{\prime\prime}$ pixels. Measurements used apertures of 4-pixel (1.5$^{\prime\prime}$). We estimate an R magnitude of 18.73 $\pm$ 0.05 for the 3$^{\prime\prime}$ contaminator, providing 1$\%$ of the flux measured in the \emph{Kepler} aperture. We also identify an additional faint source located northeast of KIC 4862625. This star is 20.65 $\pm$ 0.09 R magnitude, much fainter than the 3$^{\prime\prime}$ contaminator and contributes well less than a percent flux dilution in the \emph{Kepler} light curve. Any additional sources present within the \emph{Kepler} photometric aperture are fainter than 21st magnitude and will have negligible impact on the \emph{Kepler} light curve for KIC 4862625. 

The dominant sources of additional flux to KIC 4862625's light curve are the 0.7$^{\prime\prime}$ and 3$^{\prime\prime}$ contaminators. Combining the contributions from both stars, we assume 12-13$\%$ flux contamination to the \emph{Kepler} photometric aperture. Accounting for the dilution of the transit, the flux drop caused by the planet now corrected to be 0.1$\%$. We account for this additional light when modeling the light curve and assessing PH1b's properties (discussed in Section \ref{section:photo}). In Section \ref{sec:deconvoultion}, we revise this estimate assuming the stellar properties obtained from the photometric-dynamical  model and stellar evolution models. 
 
\section{Radial Velocity Observations}
\label{sec:rvs}

Radial velocities were obtained with the HIgh Resolution Echelle Spectrometer (HIRES) \citep{1994SPIE.2198..362V} on the Keck I telescope on Mauna Kea, Hawaii from 2012 April 1- August 4. We note that the time baseline and precision of these radial velocity measurements are insufficient to constrain the mass of PH1b and only characterize the properties of the close binary (Aa and Ab). These observations were taken with the standard setup used by the California Planet Survey (CPS) \citep{2010ApJ...721.1467H, 2012arXiv1207.6212C} with a typical resolution of R $\sim$55,000 covering 3642 to 7990 \AA. All Keck observations were taken with the C2 decker, except for the first observation, which was taken with the B5 decker. The observations were reduced using a standard reduction and extraction pipeline developed for the CPS \citep{2010ApJ...721.1467H, 2012arXiv1207.6212C} and then corrected for cosmic rays. 

The radial velocities were measured using the spectral line broadening function (BF) technique \citep{1992AJ....104.1968R}. This method lends itself particularly well to double lined spectroscopic binaries where the radial velocity difference between the stars is on the order of the resolution of the spectral observation or for high rotation stars where there is significant broadening of the spectral lines. To compute the broadening functions, a spectrum of a slowly rotating single star is required as a reference template. A single template spectra was used in the BF analysis for all the radial velocity observations of KIC 4862625. The HIRES spectrum of HD 169830, an F7V star, taken on 2012 April 1 UT was used for this purpose in the BF analysis. The radial velocity of the template star was assumed to be -17.4 km s$^{-1}$. The analysis returns a BF defined over a velocity range for each set of spectral lines present in the observation. For each component, the respective broadening function is fit to an analytic broadening kernel and the peak value is taken as the measured radial velocity. A barycentric correction is applied to each of the obtained velocities and the contribution of the template radial velocity is removed to obtain the true Doppler shifts for each component identified in the BF analysis. 

\begin{figure}
\epsscale{1.0}
\plotone{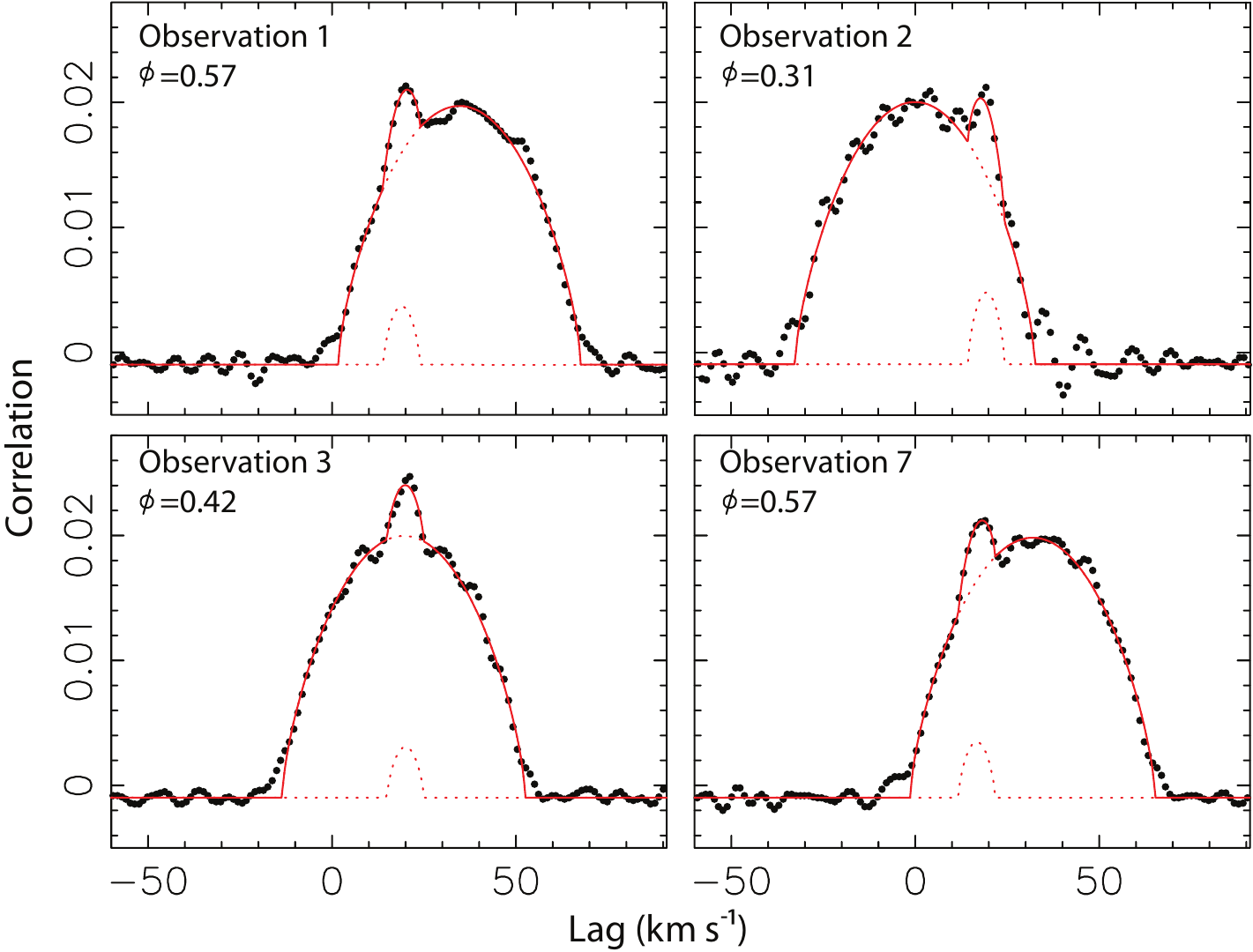}
\caption{The Broadening functions and the analytic broadening kernel fits (solid line for the sum, dotted lines for each component separately) for a subset of the Keck HIRES radial velocity observations taken at different binary orbital phases ($\phi$).}
\label{fig:BF}
\end{figure}

 HIRES spectra are split on three CCD chips: blue (3642.9-4794.5 \AA), middle (4976.6-6421.4 \AA), and red (6543.2-7989.8 \AA). The BF analysis is applied separately to the spectra obtained on each chip. The majority of HIRES observations were taken with the iodine cell in, which introduces the iodine spectra superimposed on the stellar spectra on the middle chip. We find the HIRES spectra are double-lined with two BFs: one with a narrow width and a second much broader function. Due to the presence of the second velocity component, we chose to restrict ourselves to the blue chip to measure radial velocities because the iodine lines made it more difficult to fit the BFs on the middle chip, and the BFs on the red chip were weaker than those obtained on the blue chip. A representative sample of the BFs obtained from the blue chip is plotted in Figure \ref{fig:BF}. 

\begin{figure}
\epsscale{0.75}
\plotone{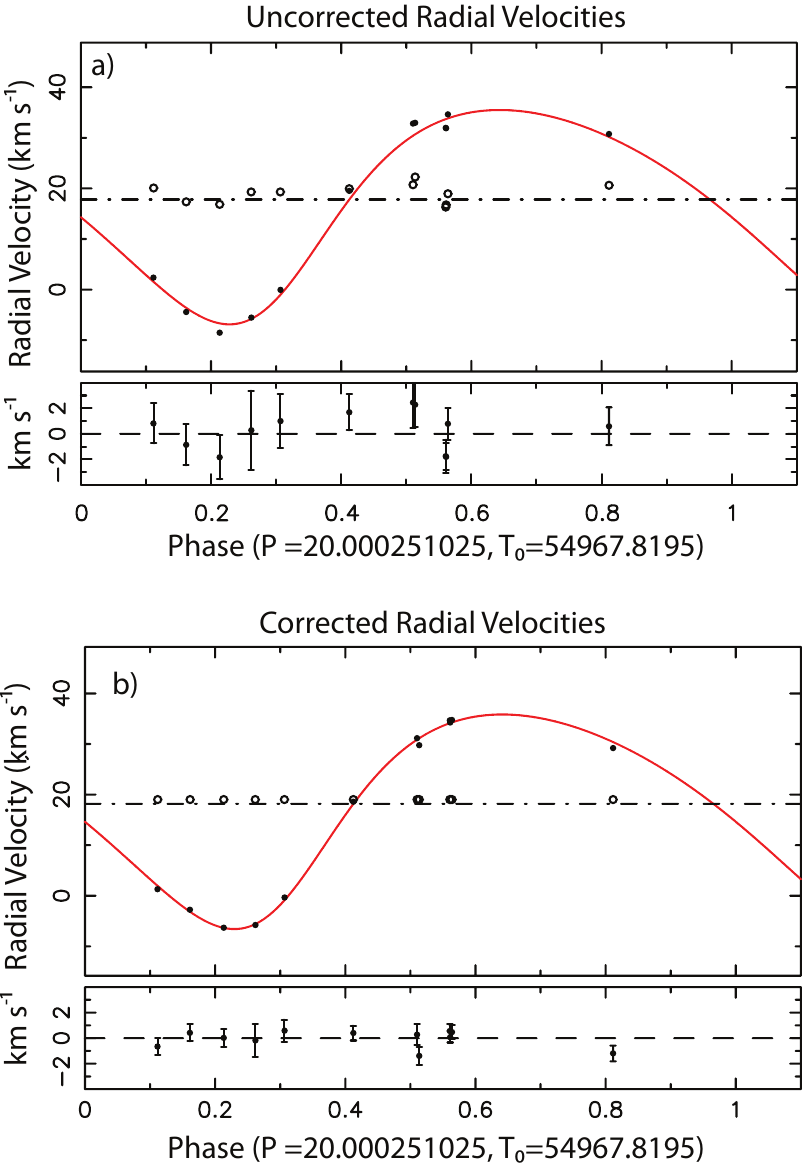}
\caption{a) Top: Uncorrected radial velocities obtained for KIC 4862625 and ELC fit to the light curve and radial velocity observations. Bottom:  The residuals of the ELC fit.  b)  Top: Offset-corrected radial velocities obtained for KIC 4862625 and improved ELC fit to the light curve and radial velocity observations. Bottom:  The residuals of the ELC fit.  - In both plots: The filled dots are the measurements for KIC 4862625. The circles are the radial velocities of the narrow broadening function component, the 0.7$^{\prime\prime}$ contaminator. Solid line (red in online version) is the best-fitting ELC solution.}
\label{fig:RVfit}
\end{figure}

Table \ref{tab:uncorrectedRVs} lists the measured radial velocities and uncertainties for both BFs, and Figure \ref{fig:RVfit}a plots the radial velocities obtained from both the wide and narrow BFs. A two component model was used for all observations except for observations 9 and 12, which due to a nearly full moon, required an additional third component in order to account for the contribution from sky line emission. The peak velocity of the wide component varies as a function of the binary orbital period and is the contribution from KIC 4862625. The narrow component is due to another source and is unchanging with a nearly constant radial velocity over the 5 month time baseline of the HIRES observations. The C2 decker is 14.0$^{\prime\prime}$ long and 0.861$^{\prime\prime}$ wide. The B5 decker is the same width but only 3.5$^{\prime\prime}$ long. For all observations the HIRES slit was aligned East-West on the sky. Located at a position angle of 123 degrees East of North, the 0.7$^{\prime\prime}$ contaminator contributes light into the HIRES slit at the $\sim$10$\%$ level, and thus is the additional narrow component identified in the BF analysis. 

The radial velocities were fit with the Eclipsing Light Curve (ELC) code \citep{2000A&A...364..265O}. Although the spectra are double-lined the close binary (Aa and Ab) is a single-lined spectroscopic binary (SB1). Star Ab is too faint to be detected in the radial velocity observations, and therefore we do not directly obtain the independent masses of Aa and Ab from the HIRES observations alone. The ELC modeling ignores the gravitational effects of the planet and assumes a two-body system. The ELC modeling cannot constrain the mass of the binary. Instead with the radial velocity observations, the ELC fits for the binary's (Aa and Ab) eccentricity ($e$), inclination ($i$), orbital period (P), time of conjunction (T$_{\rm{conj}}$), argument of periastron ($w$), radial velocity semi-amplitude ($K_A$). The \emph{Kepler} light curve was used in the fit to force a consistent binary orbital phase and eccentricity. The best-fit parameters are reported in Table \ref{tab:rvfit}. The resulting radial velocity model and radial velocity residuals are plotted in Figure \ref{fig:RVfit}a. The radial velocity measurement error for KIC 4862625 was scaled by a factor of 11.2 to produce a reduced $\chi^2$ of 1 for the model fit.  We note the 0.7$^{\prime\prime}$ contaminator has a mean radial velocity (19.12 $\pm$ 0.49 km s$^{-1}$) very close to that of the systemic velocity obtained for the eclipsing binary, 17.82 $\pm$ 0.03 km s$^{-1}$.  

\begin{figure}
\epsscale{1.0}
\plotone{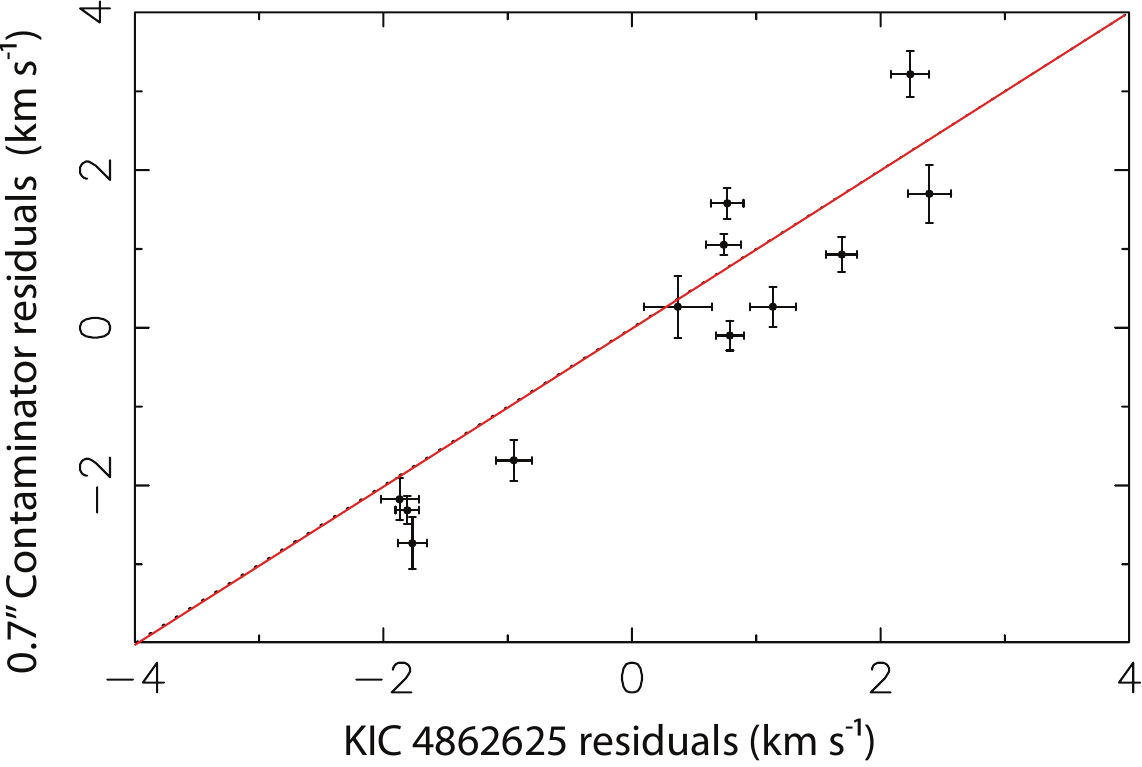}
\caption{Radial velocity residuals for KIC 4862625 versus the radial velocities residuals  of the narrow broadening function component, the 0.7$^{\prime\prime}$ contaminator before  applying the offset-correction.  The plotted line (red in the online version) is the 1 to 1 correlation in primary and companion radial velocity residuals.}
\label{fig:RVresiduals}
\end{figure}

The ELC fit for KIC 4862625 is poor with 1-2 km s$^{-1}$ residuals for the best-fit solution. With the CPS observing setup, HIRES is aligned such that ThAr lines across the echelle format fall on the same CCD pixels, to sub-pixel precision such that the obtained HIRES spectra have the same wavelength solution, accurate to about a pixel (1.4 km  s$^{-1}$). The HIRES wavelength solution zeropoint and dispersion varies within a 1 pixel between observations within a single night and on separate nights, which is on the order of 1 km s$^{-1}$. We do not correct for this effect in the BF analysis and this is thus the largest source of uncertainty in our velocity measurements. Subtracting the mean radial velocity of  the 0.7$^{\prime\prime}$ contaminator  from its measured values, we find that the residuals are directly correlated with the  ELC fit residuals for KIC 4862625 (see Figure \ref{fig:RVresiduals}). If we assume the 0.7$^{\prime\prime}$ contaminator has a constant radial velocity, we can use it to correct for the instrumental systematics in each observation's wavelength solution. We apply the narrow BF velocity residual as a linear offset to correct each of the radial velocities measured for KIC 4862625. 

With the offset-corrected radial velocity observations, the ELC fit is significantly improved with typical residuals of 0.5 km s$^{-1}$ or smaller. The radial velocity measurement error for KIC 4862625 was scaled by a factor of 4.63 to produce a reduced $\chi^2$ of 1. Table \ref{tab:RVs} lists the offset corrected radial velocities for KIC 4862625, and the model parameters are reported in Table \ref{tab:rvfit}.The ELC model fit model and residuals for the revised KIC 4862625 radial velocities are plotted in Figure \ref{fig:RVfit}b. Overplotted on Figure \ref{fig:RVfit}b are the radial velocities of the narrow BF component, the 0.7$^{\prime\prime}$ contaminator, also corrected by the same instrumental offset applied to the KIC 4862625 values.

With the correction, the systemic radial velocity of KIC 4862625 is found to be 18.18 $\pm$ 0.03 km s$^{-1}$. The average velocity of the 0.7$^{\prime\prime}$ contaminator (19.12 $\pm$ 0.49 km s$^{-1}$) is within the 2-$\sigma$ uncertainty of the binary's systemic radial velocity. Within measurement uncertainty the 0.7$^{\prime\prime}$ contaminator has the same radial velocity as the systemic velocity of KIC 4862625. Even before the offset corrections for the radial velocity values, the measured systemic velocity of the binary was 17.82 $\pm$ 0.03 km s$^{-1}$, within 3-$\sigma$ of the mean radial velocity of the 0.7$^{\prime\prime}$ contaminator. Given that the velocity dispersion of random field stars in the Galaxy is between 20 and 60 km s$^{-1}$ \citep[e.g.][]{1977A&A....60..263W}, the radial velocity of the 0.7$^{\prime\prime}$ contaminator matches the systemic velocity of KIC 4862625. At such large separations ($\sim$1000s of AU), the barycentric motion about the center of mass  in a wide binary is a  marginal contribution to the motion of the system. For stars separated by 1000 AU or more depending on on-sky projection effects and the orientation of the orbit, the radial velocity contribution due to the orbital motion of the binary components would be on the order of a few km/s or less.  Gravitationally bound components of a wide binary should therefore have similar proper motions and systemic radial velocities. This criterion has been used to identify wide binary star systems \citep[e.g.][]{1997A&AS..124...75T, 2009AJ....137.3646P}, and we adopt it here. Coupled with the close on-sky proximity of the sources, the observational evidence supports that Aa, Ab, and the 0.7$^{\prime\prime}$ contaminator are common radial velocity stars and therefore associated. We note that follow-up observations  to measure the common proper motion or orbital acceleration of these stars in the future could  solidly confirm their association.  Thus, the third source, the 0.7$^{\prime\prime}$ contaminator, is bound to the eclipsing binary (Aa and Ab) in a hierarchical architecture. We further discuss the implications in Section \ref{sec:extrastars}. 

 \section{Additional Stellar Companions}
 \label{sec:extrastars}
 
\begin{figure}
\epsscale{1.0}
\plotone{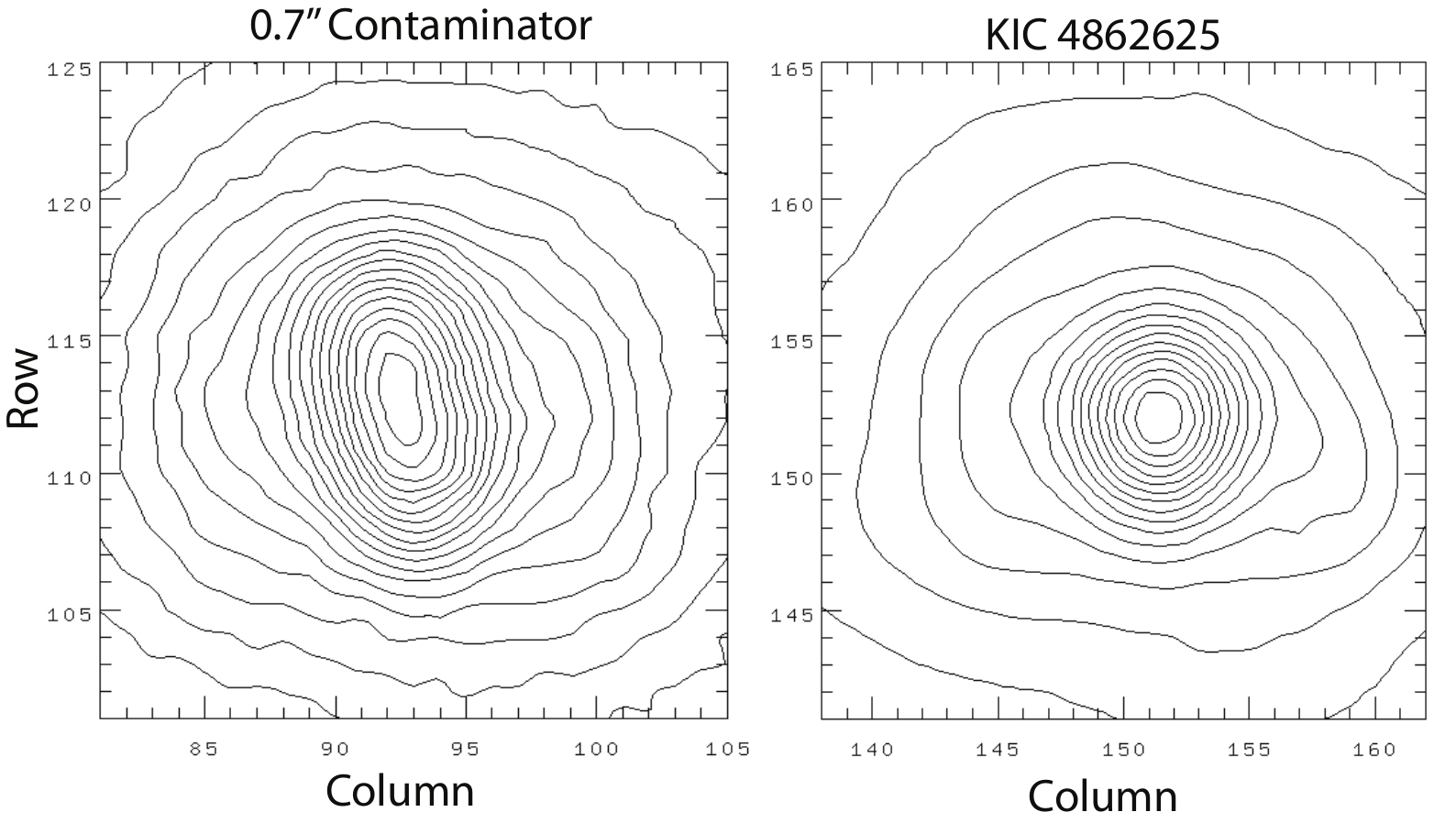}
\caption{Contour plot of  the 0.7$^{\prime\prime}$ contaminator (starting at 1049.404 counts) and KIC 4862625 (starting at 3940.448 counts) from the stacked NGS AO NIRC2 Ks observations. Contour intervals are 300 counts in both plots.}
\label{fig:contourplot}
\end{figure}

The radial velocity observations strongly indicate that  the 0.7$^{\prime\prime}$ contaminator (located at roughly $\sim$1000 AU from KIC 4862625) is likely associated with the eclipsing binary (Aa and Ab) in a wide binary.  Further inspection of the AO observations reveals that the 0.7$^{\prime\prime}$ contaminator is itself a visible double. The point-spread function (PSF) of the 0.7$^{\prime\prime}$ contaminator is elongated compared to the circular PSF of KIC 4862625 (see Figure \ref{fig:contourplot}). The one-dimensional full width-half maximum (FWHM) in the vertical direction is almost exactly twice as wide as the primary (KIC 4862625). This was observed on both nights of AO observations. Thus, the 0.7$^{\prime\prime}$ contaminator is itself a wide visual binary, making the KIC 4862625 system a quadruple star system, and the first identified to host a planet. 

The AO observations barely resolve the outer star system, which we will label `Ba' and `Bb'. We are unable to ascertain an accurate projected separation between the two stars, but we estimate that it must be  less than 0.04$^{\prime\prime}$ ( $\lesssim$40 AU).  In Section \ref{sec:deconvoultion}, using stellar evolution models,  we estimate the spectral type of Ba and Bb and more accurately estimate both their projected separation and their projected distance to KIC 4862625. Our rough estimate of the AO $K_s$-band brightness difference of $1.0 \pm 1.0$ mag. Both Ba and Bb contribute flux to the Keck spectra. Both stars contribute to the second broadening function identified in the Keck spectra. If Ba and Bb are equal brightness and had significantly different velocities we would expect to see a third broadening function. The resolving power
of the BF analysis corresponds to $\sim$5 km s$^{-1}$. If the majority of the contaminating flux is associated with Ba, the physical proximately of Bb suggests it is bound to Ba and therefore will have the same radial velocity as Ba. Thus, we claim that Ba and Bb  both have the systemic radial velocity of Aa and Ab. Therefore they are both associated with the close binary (Aa and Ab) and are likely bound gravitationally to the system.

This quadruple stellar system is configured as a close eclipsing binary (Aa and Ab) and a wide binary (Ba and Bb) arranged with a projected distance between the two binaries that is larger than the physical separation between the individual stars in the binary components. This is not an atypical stellar architecture; \cite{2001IAUS..200...84T} and \cite{2006A&A...459..909C} identify nearly half of their sample of quadruple systems in such a two pair hierarchical structure. This type of hierarchal system  may be a natural outcome of the dynamics during the dissipation of embedded clusters \citep{2010MNRAS.404.1835K}. The two binary pairs are only weakly bound to each other. Interactions with passing stars and molecular clouds can dynamically disrupt and ionize wide binaries with separations on the order of 1000 AU \citep{1986Icar...65...13H,2010MNRAS.401..977J}.  \cite{2013Natur.493..381K} find for two star wide binaries, after 10 gigayears 90$\%$ of systems with semimajor axes greater than 3000 AU are ionized. Those with smaller semimajor axes survive much longer. We expect the results to be similar for two -pair hierarchical systems, and thus expect such stellar systems separated by a few thousand AU to survive for 10 gigayears  or longer.  Thus the presence of the additional Ba and Bb does not significantly constrain the age of the planetary system.  

With the radial velocity observations we can only ascertain that the visual binary is gravitationally bound to stars Aa and Ab. With only one epoch of adaptive optics observations we do not have a sufficient time baseline, to measure their motion on sky, and accurately determine their orbit around Aa and Ab. Future adaptive optics observations may be able to measure a proper motion and constrain the orbits of the visual binary. Orbital integrations with MERCURY \citep{1999MNRAS.304..793C} of Ba and Bb assuming circular orbits with semimajor axes of 1000 AU around Aa and Ab find that on timescales of gigayears, there is little effect from the distant binary on the planet's orbit and the eclipsing binary's (Aa and Ab) orbits. This is consistent with the results of \cite{2013Natur.493..381K} who find that only for wide-binary systems, with semimajor axes less than 1000 AU, is eccentricity pumping observed in planetary orbits with orbital periods ranging from 12-165 years (5-30 AU). Thus with a much more tightly bound orbit around Aa+Ab, the dynamical effects from Ba and Bb on the planet in this system is negligible.

\section{Photometric-Dynamical Model}
\label{section:photo}

Previous analyses combining transiting circumbinary planetary light curves and radial velocity data  \citep{2011Sci...333.1602D, 2012Natur.481..475W,2012Sci...337.1511O,2012ApJ...758...87O} have demonstrated that it is possible to measure absolute masses and radii of the parent stars and planet from first principles. From the light curve, one may measure the radii ratios between stars (assuming they both eclipse) and between the stars and planet. Also from the light curve, transverse line-of-sight velocities, normalized by the primary radius, may be measured for the bodies at the conjunctions. The densities of the stars and planet are determined by these normalized velocities, eccentricities, arguments of periapse and the stellar mass ratio.  Stellar eccentricities and arguments may be measured directly from eclipse timings and eclipse durations while those quantities for the planetary orbit may constrained from transit durations observed at different phases of the planetary orbit.  

With eclipsing stellar binaries, the mass ratio between the stars may be determined from a double lined spectroscopic solution.  In this case, two mass functions may be measured from the radial velocity of both stars.  In previous examples and is the case with PH1, the radial velocity of the secondary (Ab) is not presently measurable.  However, with transiting circumbinary planets, this stellar mass ratio may be measured from the transit times of the planet alone.  In particular, the amplitude of the transit timing non-linearity is set by the maximum projected separation of the occulted star from the barycenter of the triple system.  This separation depends linearly on the stellar mass ratio. The non-linearity of the transit times can be almost entirely attributed to the approximately Keplerian motion of stars and planet. With this approximation and assuming a circular planetary orbit, it can be shown that $\Delta$TTV,  the peak-to-peak amplitude of the transit timing non-linearity (which may be measured from a plot of transit time versus stellar binary phase, see Figure \ref{fig:POC}), is defined as:
\begin{equation}
\frac{\Delta \rm TTV}{P_b}=\frac{1}{\pi} \frac{q}{(1+q)}\frac{P_A}{P_b}
\end{equation}
where $q = M_{Ab}/M_{Aa}$,  $P_A$ is the orbital period of the binary, and $P_b$ is the orbital period of the planet. The relatively small timing perturbations on top of this gross Keplerian behavior, owing to the two-body potential of the stars and the mass of the planet, may be accounted for with a direct numerical integration of Newton's equations. The non-linearity of stellar eclipse times are due to a combination of physical delays and the light-time effect due to the planet; its mass relative to stars may be measured directly from the eclipse timing variations. By observing the radial velocity of the primary star (almost entirely due to the secondary), one may measure a single mass function.  The mass of the secondary can be determined from this mass function and the mass ratio measured from transit timing.  Subsequently, we may determine the masses of the remaining bodies and their radii from the densities constrained from the light curve analysis.  

The eclipse depths is the measure the ratio of the the radii of  Ab to Aa. Similarly the ratio of the planet's radius to to the radius of primary star (Aa) is derived from the measured transit depth. Other orbital parameters can be constrained with the light curve data, in addition to the radial velocity data for the stellar binary. With radial velocities measured for Aa and no constraints for the radial velocity of Ab, we can only measure of the mass function of the binary  and not the individual masses of the two stars. The mass ratio of the binary can be calculated using the planet transit observations. For a circular coplanar planetary orbit, the peak-to-peak amplitude of the timing non-linearity of the planet transits is dependent only on the mass ratio of the stars and the  ratio of the period of the stellar binary to the planet's period. With the stellar mass ratio, from the light curve data alone, and the radial velocity curve,  the density of the primary star (Aa) can be found, thus providing radii estimates for all three bodies. Three body non-Keplerlian dynamics observed in the light curve then enable the mass of the planet to be constrained. The variations in the timing of the primary and secondary can be induced by the planet depending on its mass. If no eclipse timing offsets are measured, as in our case, only an upper limit on the planet mass can be calculated. 

To constrain the orbital and physical properties of the planetary system, we employ a `photometric-dynamical' model, as has been applied to the previously discovered \emph{Kepler} transiting circumbinary planets \citep{2011Sci...333.1602D, 2012Natur.481..475W,2012Sci...337.1511O,2012ApJ...758...87O} simultaneously fitting the obtained radial velocity observations and \emph{Kepler} light curve data.  In our analysis, we only estimate the orbital and physical properties of Aa and Ab and the transiting planet. The effects of the wide binary (Ba and Bb) are not observable in the Kepler data, and thus were not included in our dynamical model. In this model, the two host stars (Aa and Ab) and the planet interact according to Newtonian gravity. The initial coordinates of the three bodies (Aa, Ab, and the planet) are specified at a particular time (in this case $t_0 =$ 2,454,970 BJD) and their positions, as well as the radial velocity of Aa, at any other time are calculated via numerical integration. Positions are calculated at the times of the {\em Kepler} observations and used as inputs to produce a light curve model. The light curve model presumes spherical planet and stars with radial brightness profiles specified according to a quadratic limb darkening law ($I(\mu)=I_0(1-u_{Aa}(1-\mu)-v_{Aa}(1-\mu)^2)$). The model is computed for  time steps shorter than \emph{Kepler}'s  29.4244 minute observations. To account for the finite exposure time we use gaussian-quadrature to integrate the continuous model in order to directly compare to the observational data.  For a more detailed description of the photometric-dynamical model, we refer the reader to \cite{2011Sci...333.1602D}.  For this specific circumbinary planetary system, the photometric-dynamical model describing all stellar eclipse and planet transit events has 24 parameters discussed further below. These fit parameters were  chosen to minimize correlations between parameters, and all have uniform priors (over open intervals unless otherwise explicitly stated). 

Eleven parameters describe the Jacobian coordinates of the stellar binary (Aa and Ab) and the binary comprised of the planet and  center-of-mass of the Aa+Ab system. These are osculating Keplerian terms (one set for the planet's orbit around the center-of-mass of Aa+Ab and one set describing the orbit of the stellar binary) that are slowly varying in time. Each binary is described by a period, two linearly-independent parameters involving the eccentricities and arguments of periastron, a sky-plane inclination, a time of conjunction and longitude of the ascending node The absolute nodal longitude (relative to North) of the system cannot be determined from the data alone; we therefore choose to fix the longitude of the stellar binary to zero.  We encode the eccentricity and argument of periastron of the stellar binary as components of a vectorial eccentricity: $\vec{e_A} = [e_A \sin \omega_A, e_A \cos \omega_A]$.  Those for the binary including the planet involve a square-root in the amplitudes ($[\sqrt{e_b} \sin \omega_b, \sqrt{e_b} \cos \omega_b]$) to simplify the specification of uniform priors in $e_b$ and $\omega_b$.  We note that deviations from the sinusoidal expectation of the planet's O-C variations as a function of  binary phase (see  Figure \ref{fig:POC}) constrain the eccentricity of the planet and binary in addition to the orbital angles of the planet.

We must also specify the masses of all three bodies in our photometric-dynamical framework. We do so by fitting for the mass of star Aa (times the gravitational constant), $G M_{Aa}$, and for the masses of Ab ($M_{Ab}$) and the planet ($M_{b}$) we solve for the mass ratios relative to $M_{Aa}$: $q = M_{Ab}/M_{Aa}$ and $q_p = M_{b}/M_{Aa}$ - both of which have uniform priors for positive values.  We also parameterize the density of star Aa-- this quantity is directly measurable from the light curve when the mass ratio $q$ is constrained from the planetary transits and the transit timing deviation from a  Keplerian orbit about the stellar binary's center of mass. With no perceptible eclipse timing variations in the \emph{Kepler} light curve (see Section \ref{sec:lightcurve}) induced by the planet, the model fit will only be able to provide an upper limit on $q_p$ as well as the derived value for the planet's mass. 

Six more model parameters describe the eclipse and transit light curve: the radius ratios of stars and planet ($R_{Ab}/R_{Aa}$, $R_{b}/R_{Aa}$ respectively) the contaminating flux in the Kepler bandpass relative to that of star Aa ($F_{X}/F_{Aa}$), the flux of star Ab to Aa ($F_{Ab}/F_{Aa}$), and two limb darkening parameters of star Aa ($u_{Aa}$ and $v_{Aa}$). The limb darkening parameters were restricted to physically plausible values with the linear constraints $0< u_{Aa} < 1$, $u_{Aa}+v_{Aa} < 1$. The limb darkening coefficients of the secondary star had a negligible affect of the overall fit and were fixed to $u_{Ab} = 0.6$, $v_{Ab} = 0.1$.  The width of the assumed normal, uncorrelated noise of the photometric data was also fitted ($\sigma_{LC}$). Lastly we specify the offset of the radial velocity data and a radial velocity jitter term, added in quadrature to the formal radial velocity errors.

The raw \emph{Kepler} SAP light curve data for the individual planet transits and stellar eclipses including observations 1 day before and after the midpoint of each event are extracted and used in this analysis. This corresponds to 5302 cadences of \emph{Kepler} data at primary or secondary eclipse and 521 cadences of data near planetary transit (some of these cadences appear in both data sets as transits occur in close proximity to eclipses, at times).  The isolated planet transits and stellar eclipses were initially corrected for a cubic trend in time.  After an initial photometric-dynamical fit, this cubic fit was repeated, this time including all data after having divided out this best-fitting model for the eclipses. The level of the out-of-eclipse flux is determined inline prior to the computation of the likelihood (given below).  In other words, for a given parameter set $\vec{p}$, we divide out the eclipse model then determine the mean of the residuals and divide through by this value prior to computing the likelihood.  This linear parameter is weakly correlated with the remaining parameters and trivially marginalized over.

The likelihood of a given parameter  set $\vec{p}$ was specified as:
\begin{eqnarray}
		{\cal L}(\vec{p}) &\propto&  \prod_{i} \sigma_{\rm LC}^{-1} \exp \left[-\frac{(\Delta F^{LC}_i)^2}{2 \sigma_{\rm LC}^2} \right] \times \prod_{j}^{N_{\rm RV}} \left(\sigma_{j}^2+\sigma_{RV}^2\right)^{-1/2} \exp \left[ -\frac{(\Delta V_j)^2}{2\left(\sigma_{j}^2+\sigma_{RV}^2\right)} \right] \\ \nonumber 
		&& \times \exp\left[-\frac{\left(F_X/F_{Aa} - 0.12\right)^2}{2 (0.02)^2}\right]\label{eqn:like}
\end{eqnarray}
where $\Delta F^{\rm LC}_i$ is the $i$th photometric data residual, $\sigma_{\rm LC}$ is the width parameter describing the photometric noise of the long cadence data, $\Delta V_j$ is the $j$th radial velocity residual, $\sigma_j$ is the uncertainty in the $j$th radial velocity measurement and $\sigma_{RV}$ is the stellar jitter term added in quadrature with the $\sigma_j$. The final product in the second line of the above equation enforces a Gaussian prior on the contamination of $F_X/F_{Aa} = 0.12\pm0.02$, appropriate given the discussion in Section \ref{sec:contamination}. 

\begin{figure}
\epsscale{1.0}
\plotone{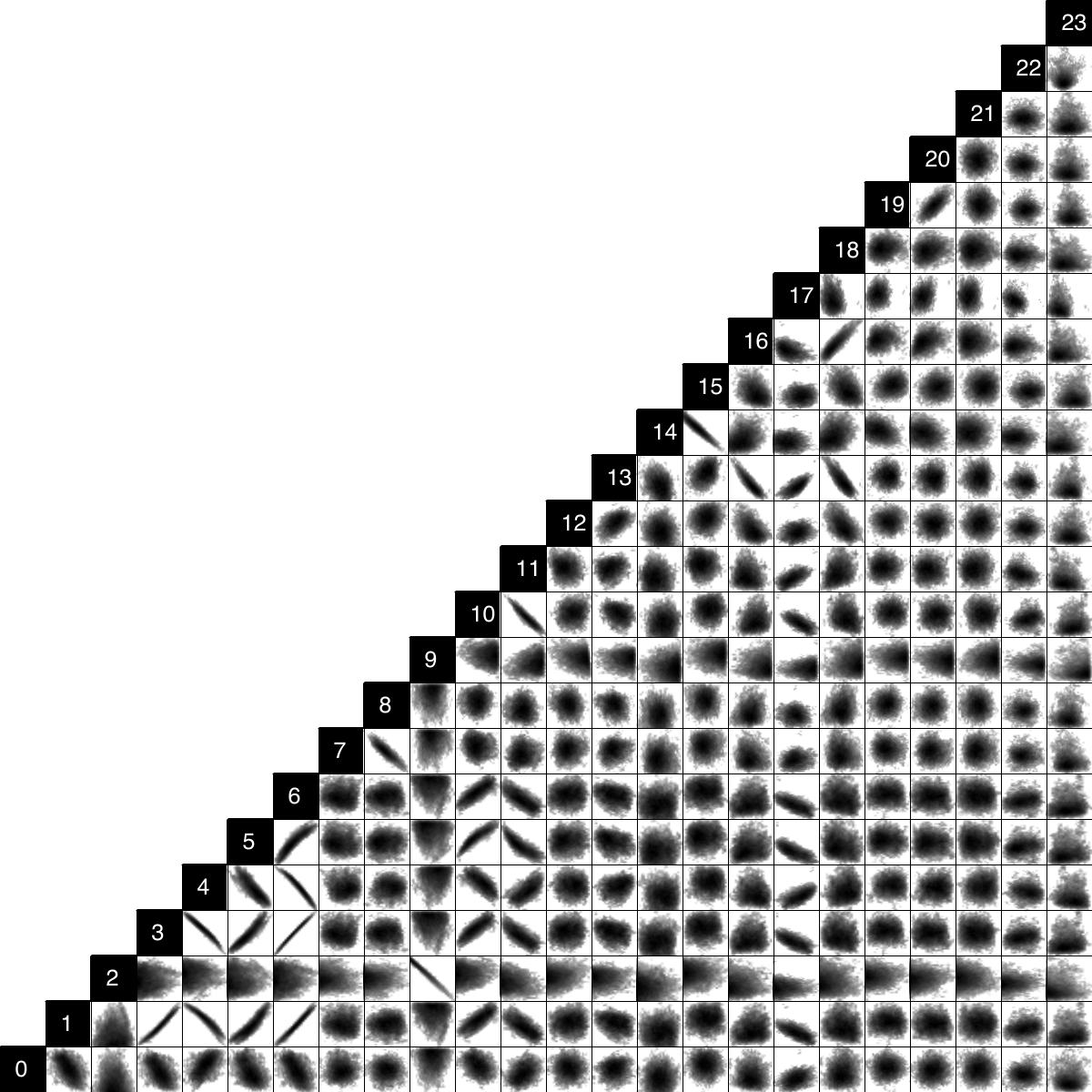}
\caption{Two-parameter joint posterior distributions for the best-fitting  'joint' solution model parameters. The 68$\%$ and  95$\%$ confidence regions are plotted logarithmically in order to elucidate the nature of parameter correlations. The indices listed along the diagonal indicate which parameter is associated with the corresponding row and column. The parameter name corresponds to index  value listed in Table \ref{tab:Parameters}. }
\label{fig:correlation}
\end{figure}

The best-fitting model was obtained by maximizing the likelihood ${\cal L}(\vec{p})$ for each of the 24 parameters. We first performed a fit to the combined eclipse and planetary transit data (hereafter referred to as the `joint solution') and determined credible intervals in the above parameters using a Differential Evolution Markov Chain Monte Carlo algorithm (DE-MCMC; ted Braak 2008). We generated a population of 100 Markov chains in an overdispersed region covering the final posterior and evolved through 19,000 generations until the chains showed adequate convergence according to the standard metrics. These chains were concatenated after removing a portion of `burn-in.'  Results from this analysis are listed in the first data column of Table~\ref{tab:Parameters} with additional derived physical and orbital parameters reported in the first column of Table~\ref{tab:DerivedParameters} The values reported for each fit parameter are the median of the posterior distribution and the $68\%$ confidence level.  To examine the degeneracy of the fit solution, we provide a correlation plot in Figure \ref{fig:correlation} comparing each of the fit parameters. We find only weak nonlinear correlations exist between the fit parameters.

As a self consistency check of our best-fit solution, we isolated the data involving the stellar eclipses alone (hereafter referred to as the `EB-only' solution) and repeated our analysis with the same photometric-dynamical model.  We varied only those parameters relevant to modeling the eclipsing binary, holding the mass of A constant. We force the planetary mass ratio to be zero and fixed all osculating Keplerian elements related to the planet. Thus, the system modeled reduces to a Keplerian two-body problem.  Without the timings of the planetary events, we cannot constrain the stellar mass ratio and are unable to provide absolute dimensions to the stars. We choose therefore to report from this analysis standard quantities for of an eclipsing, single-lined spectroscopic binary (SB1). Fifty chains were evolved through 25,000 generations in the DE-MCMC. The best-fit results  from the EB-only solution  and subsequent parameter posterior calculation via Markov Chain Monte Carlo simulation are provided in the second data column of Table~\ref{tab:Parameters} and of Table~\ref{tab:DerivedParameters}.  As expected, we find the EB-only analysis has no preference in contamination -- its marginalized posterior reflects the prior.

In previous circumbinary planetary systems modeled with this photometric-dynamical approach  \citep{2011Sci...333.1602D, 2012Natur.481..475W,2012Sci...337.1511O,2012ApJ...758...87O}, the eclipses of the stellar binary and the transit events due to the planet(s) were modeled self-consistently.  In our case, the EB-only solution (fitting the only the stellar binary properties) is preferred over the joint solution (fitting both the stellar binary and planet ) with $\Delta \chi^2 \approx 30$ when considering just the data involving the stellar eclipses, the radial velocity data and the contamination prior. This error is barely perceptible by eye in the photometric residuals and is only weakly significant in the resulting parameter posteriors; however, the bias is statistically clear. Investigating the parameters from the joint solution, we see that the contamination term is $2 \sigma$ lower than the expectation (even with a Gaussian prior enforced). The remaining EB-only parameters, which are correlated with the contamination level, are also $1-2\sigma$ biased relative to the joint solution equivalent parameters as a consequence. 
  
The {\em joint} solution  yielded  $1-2 \sigma$ departures from those same parameters estimated in the EB-only solution. We suspect that our inability to accurately model the planetary events in the joint-solution is driving the bias in contamination. This systematic  is only marginally detected, making isolating the root cause difficult. The failure of the joint model to adequately describe the data may be due to either missing mass components (e.g., a fourth body), an overly simplified treatment of the light curve noise  (e.g., correlated noise may be corrupting information gleaned from the planetary transits, ) and/or deterministic light curve features that could not be clearly modeled in this analysis (spot crossing during transits). Given the significance of the anomaly, rather than invoking a fourth body, we believe the latter two options are the more likely root cause. The EB-only model is less sensitive to potential correlated noise sources (stochastic sources or from spot-crossings, for example), as the stellar eclipse events are approximately identical at each epoch (outside of insignificant timing anomalies due to the planet), and provides a more robust measure of the uncorrelated noise component. 

\begin{figure}
\epsscale{1.0}
\plotone{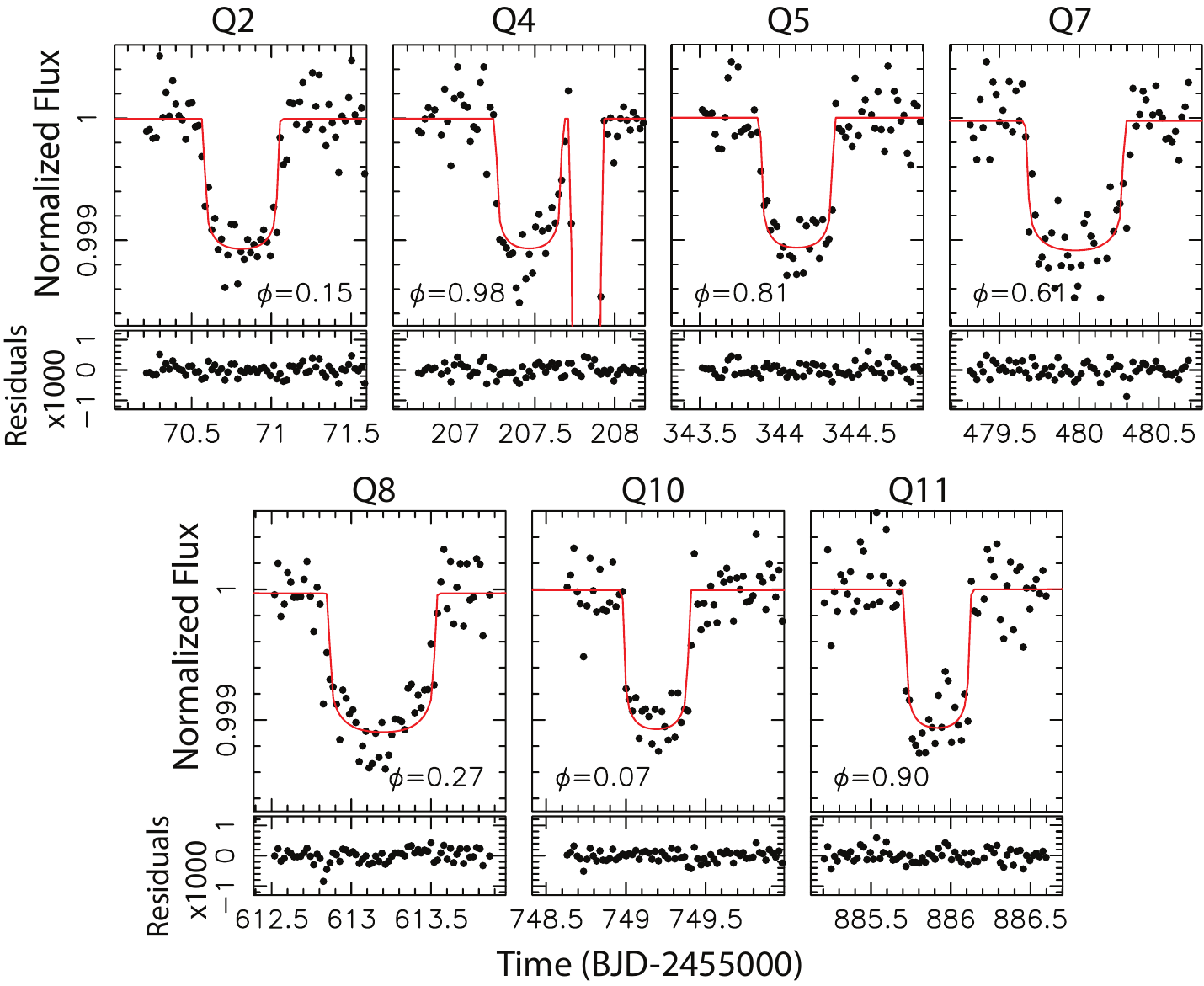}
\caption{Best-fitting EB-fixed photometric-dynamical model ( solid line -red in the online version) and reduced \emph{Kepler} light curve isolated for the planet transits. Fit residuals for the model compared to the light curve data are also shown below  each transit plot. In Quarter 4 (Q4) , the planet transit occurs just before a primary eclipse.  }
\label{fig:photofitlc}. 
\end{figure}

Given that the EB-only solution is less sensitive to potential correlated noise sources, we compute a third analysis (referred to hereafter as the `EB-fixed' solution) in order to robustly estimate the planet parameters. In the EB-fixed analysis  we use the stellar values found in the EB-only solution fitting for the planetary binary orbit and the stellar mass ratio considering only data involving the planetary transits.  In practice, this involved fixing the 11 osculating elements of the stellar binary, the normalized lengths $R_{Aa}/a$, $R_{Ab}/R_{Aa}$, flux ratios, limb darkening coefficients and the mass function as determined from the best-fitting EB-only model. Forty chains were evolved through 64,000 generations in the DE-MCMC. The results of this analysis are reported in the final data column of Table~\ref{tab:Parameters} with additional derived physical and orbital parameters reported in the last column of Table~\ref{tab:DerivedParameters}.  In addition to those parameters directly measured from this analysis (the osculating elements of the planetary binary, the stellar mass ratio and the planetary radius ratio) we report parameter posteriors (daggered values in Table~\ref{tab:Parameters}) assuming the independence of EB-only and EB-fixed solutions. For example, the mass of star Aa is determined assuming the mass function determined in the EB-only analysis and the mass ratio in the EB-fixed analysis are independent (simple propagation of errors renders our resulting hybrid posterior). Figure \ref{fig:photofitlc} plots the reduced \emph{Kepler} photometric planet transit observations used in the photometric-dynamical analysis , the best-fitting  EB-fixed solution model, and the residuals between the observed light curve and the best-fitting model. 

Adopting a fixed width for the assumed normal, uncorrelated noise of the photometric data ($\sigma_{LC}$), we can compute the traditional $\chi^2$ statistic for the various contributions to the likelihood and compare the best-fitting solutions from the three analyses described above (joint, EB-only, and EB-fixed). We adopt the value of $\sigma_{LC} = 203$ ppm as derived from the EB-only analysis. This value of $\sigma_{LC}$ is within 12\% of the root-mean-square of the best-fit EB-only residuals.  The difference in $\chi^2$ between the joint solution and the EB-only solution considering only the radial velocity is negligible.  For the 5302 cadences involving only the stellar eclipses we find $\chi_{EB}^2 = 5314$ and $\chi_{EB}^2 = 5290$ for the joint and EB-only/EB-fixed solutions, respectively. The eclipsing binary solution has 5290 degrees of freedom (after having excluded parameters relating to the RV data, the planet, the planet orbit and stellar mass ratio) which suggests the EB-only solution is marginally preferred over that from the joint solution.  When considering the 521 cadences associated with planetary transits alone, we find $\chi_{PL}^2 = 478$ and $\chi_{PL}^2 = 521$, for the  joint and EB-fixed solutions, respectively.  The model for the planetary events alone has 512 degrees of freedom. Despite the preference for the lower $\chi_{PL}^2$ from the joint solution (which accounts for the overall lower $\chi^2$ of the joint solution despite the contamination prior and the worse-fitting EB solution), the $\chi^2_{PL}$ of the EB-fixed solution is acceptable (with a 38\% probability of having a $\chi_{EB}^2$ larger than 521). Given our confidence in our estimation of the contamination prior, we conclude that the preferred EB-only solutions, tabulated in the final two columns of Table~\ref{tab:Parameters} provide the most accurate assessment of the planetary system's parameters (both for the stellar binary and planet). We adopt, however, the limit on the mass of the planet from the joint analysis. 
 
\section {ELC Model Fit}
\label{ELCmodel}
The ELC modeling can be used as an independent check on the binary properties derived from the photometric-dynamical  modeling effort. The ELC ignores the gravitational effects of the planet, which in this case are negligible over the duration of the \emph{Kepler} light curves. Because Ab is too faint to be observed in the HIRES spectrum, ELC cannot obtain an independent mass for Aa and Ab or the absolute radii of Aa and Ab. Combining the radial velocity observations and the \emph{Kepler} light curve, the ELC fits for binary's (Aa and Ab) eccentricity ($e$), inclination ($i$), orbital Period (P), time of conjunction (T$_{\rm{conj}}$), argument of perihelion ($w$), radial velocity semi-amplitude ($K_{A}$). the fractional radii of Aa and Ab (R$_{Aa}$/$a$ and R$_{Ab}$/$a$), flux contamination, the binary temperature ratio (T$_{Ab}$/T$_{Aa}$), and limb darkening parameters for star Aa  ($u_{Aa}$ and $v_{Aa}$) assuming the same quadratic limb darkening law identical to that used in the photometric-dynamical  model (as described above in Section \ref{section:photo}). Table \ref{tab:ELC} lists the physical and orbital parameters for Aa and Ab obtained from the ELC fit. The ELC-derived values are consistent with those obtained from the photometric-dynamical  modeling EB-only solution (see Table \ref{tab:Parameters}) within the $1-\sigma$ uncertainties. Thus we are confident in the physical and orbital parameters derived in the EB-fixed solution for PH1b and its host stars. 

\section{Stellar Properties}

In the following section, we further discuss and explore the stellar properties of all four stars in the hierarchical system. The resulting best-fitting parameters and their uncertainties for Aa and Ab from the photometric-dynamical  model are summarized in Table \ref{tab:Parameters}. Derived physical and orbital parameters for Aa and Ab are listed in Table \ref{tab:DerivedParameters} for the EB-fixed solution. The primary star (Aa) is an F dwarf (1.734 $\pm$  0.044 R$_\odot$,  1.528 $\pm$ 0.087 M$_\odot$) and Ab is M dwarf (0.378 $\pm$ 0.023 R$_\odot$, 0.408 $\pm$ 0.024 M$_\odot$). Using HIRES spectroscopic observations we further constrain the properties of the primary star in the eclipsing binary (Aa). With stellar evolution models, we estimate the ages and properties of the planet host stars (Aa and Ab) and the visual binary (Ba and Bb). 
 
 \subsection{High Resolution Spectroscopy}
 
To refine the physical properties of star Aa,  the HIRES spectrum  from 2012 August 4 UT,  was fit with Spectroscopy Made Easy (SME) \citep{1996A&AS..118..595V,2005ApJS..159..141V}. Modeling the stellar parameters with specific gravity remaining fixed at the value obtained from the photometric-dynamical  model, log(g)=4.14, the SME analysis finds: T$_{eff}$=6407 $\pm$ 150 K, $v$sin$i$=32.6 $\pm$ 2.0 km s$^{-1}$, and [Fe/H]= +0.21 $\pm$ 0.08 for Aa. Although Ba and Bb will have a $\sim$10$\%$ flux contribution to the spectrum, the high rotational velocity, which produces a significant broadening of the spectral lines, will be the largest source of uncertainty and constraint on the precision of the best-fit SME stellar parameters. For comparison the BF analysis which is not biased by the contaminating binary yields a rotational velocity of 32.38 $\pm$ 0.47 km s$^{-1}$.  The temperature estimate is consistent with that of an F star. Combining the SME temperature estimate of Aa with the ELC binary temperature ratio (T$_{Ab}$/T$_{Aa}$), we find Ab has a temperature of 3561 $\pm$ 150 K. 
 
 \subsection{Stellar evolution modeling}
\label{sec:stellarevolution}

Combining the results from SME, ELC, and the photometric-dynamical modeling, our determination of the masses, radii, and temperatures of both stars in the eclipsing binary (Aa and Ab), in addition to our knowledge of the metallicity of the system, permit a more stringent comparison with models of stellar evolution with fewer free parameters than if the
composition were unknown \citep[see, e.g.,][]{1991A&ARv...3...91A, 2010A&ARv..18...67T}. This test is particularly interesting for the low-mass secondary. Relatively few systems containing M dwarfs with precise (and accurate) determinations are available, and these stars have shown significant disagreements with theory in the sense that they tend to be larger and cooler than predicted \citep{2003A&A...398..239R,2002ApJ...567.1140T, 2005ApJ...631.1120L,2009ApJ...691.1400M,2012arXiv1209.1279T}

\begin{figure}
\epsscale{1.0}
\plotone{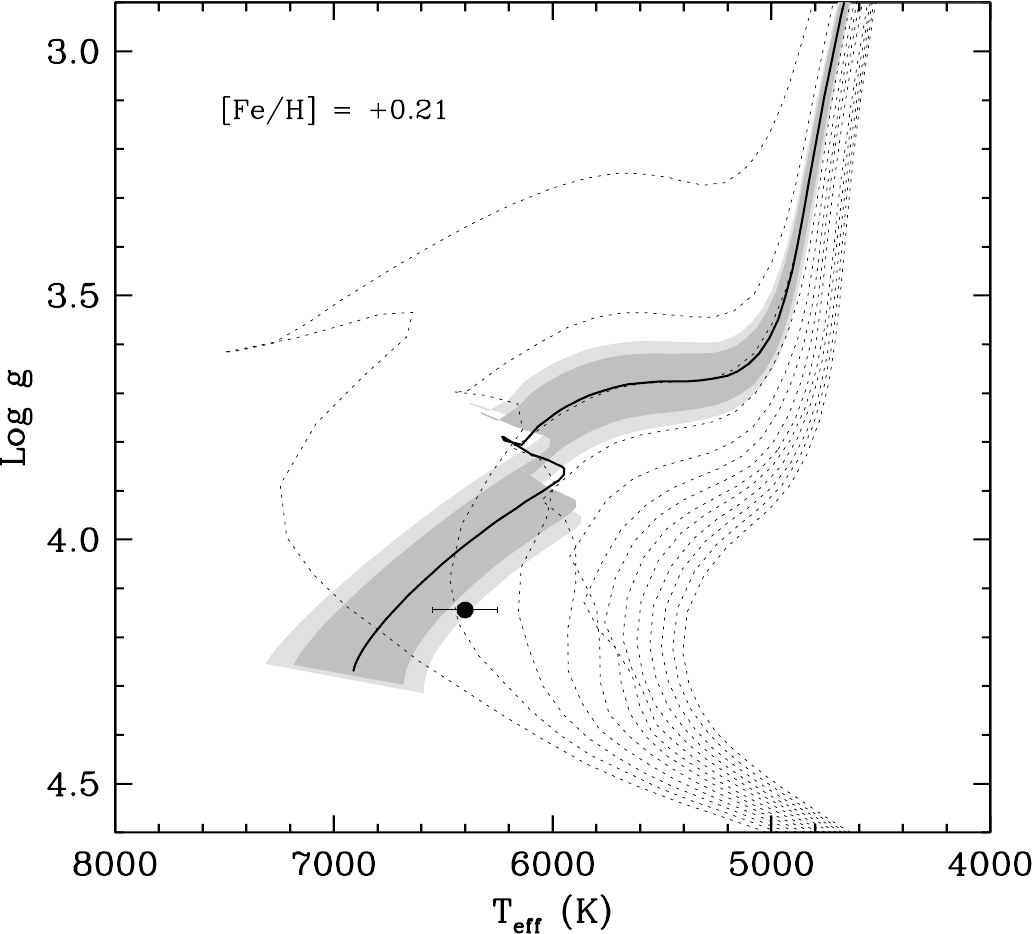}
\caption{Stellar evolution models from the Y$^2$ (Yonsei-Yale) series by
\cite{2001ApJS..136..417Y} compared against the measurements for the Aa. 
The solid line represents an evolutionary track for the measured
mass ($M_{\rm Aa} = 1.528 \pm 0.087~M_{\sun}$) and metallicity 
(${\rm [Fe/H]} = +0.21 \pm 0.08$), and the dotted lines are 1--13 Gyr
isochrones (left to right) in steps of 1 Gyr. The uncertainty in the
location of the track that comes from the observational errors is
shown by the shaded area. The darker area reflects the mass error, and
the lighter area includes also the uncertainty in the
metallicity.}
\label{fig:yale}
\end{figure}

Figure~\ref{fig:yale} compares the measured properties of the Aa star in the $T_{\rm eff}$--$\log g$ plane with models from the Y$^2$ (Yonsei-Yale) series \citep{2001ApJS..136..417Y}. An evolutionary track (solid line) is shown for the measured mass and metallicity of the star, with the 1-$\sigma$ uncertainty in the location of the track coming from the observational
errors being represented by the shaded area (dark gray reflecting the mass error only, and light gray including also the uncertainty in the metallicity).  Isochrones are indicated with dotted lines for ages from 1 to 13 Gyr. There is good simultaneous agreement with the mass, radius, and temperature of the Aa within the uncertainties, for
an age of roughly 2 Gyr, according to these models.

\begin{figure}
\epsscale{0.8}
\plotone{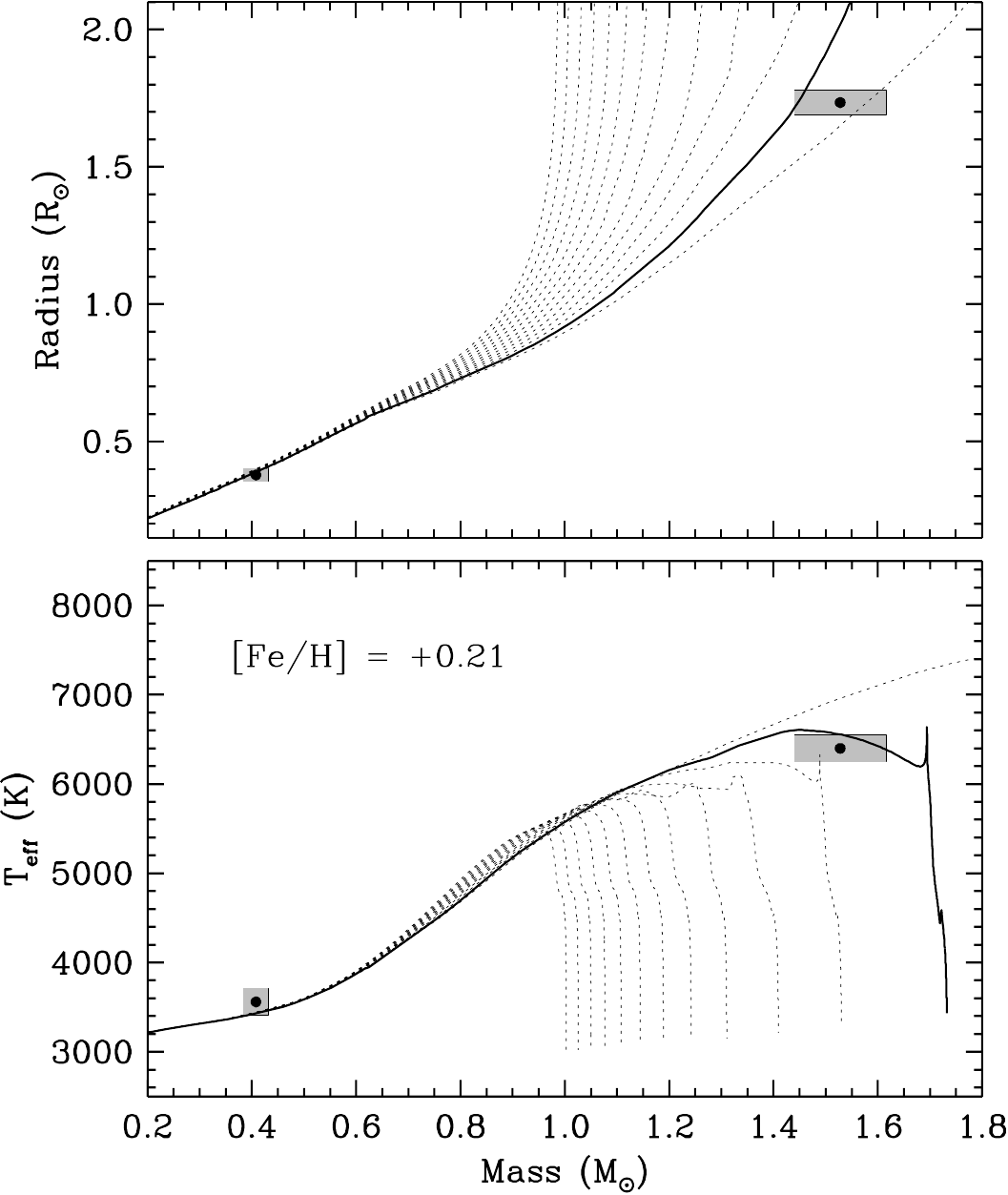}
\caption{Mass, radius, and effective temperatures for Aa and Ab
compared against 1--13 Gyr model isochrones from the Dartmouth series
by \cite{2008ApJS..178...89D}, for the measured metallicity of the system
(${\rm [Fe/H]} = +0.21 \pm 0.08$).  The solid line corresponds to a 2 Gyr
isochrone, for reference. (Top) Mass--radius diagram showing good
agreement for both stars (Aa and Ab) for an age between 1 and 2 Gyr; (Bottom)
Mass--temperature diagram again showing that the models match our
estimated temperatures, within the errors.}
\label{fig:dartmouth}
\end{figure}

In Figure~\ref{fig:dartmouth} we present another comparison using models from the Dartmouth series \citep{2008ApJS..178...89D}, in which the physical ingredients (including the equation of state and the non-gray boundary conditions between the interior and the atmosphere) are more realistic for stars significantly below 1\,$M_{\sun}$. With
the reasonable assumption that the composition is the same for the two objects, there is again satisfactory agreement between the models and the observations in the mass--radius diagram shown in the top panel, also for Ab. The age we infer based on the primary properties is 1--2 Gyr, similar to the previous estimate. The predicted flux ratio between the stars in the $K\!p$ band is 0.0011, according to these models, which is not far from our measurement based on the photometric-dynamical model (Table \ref{tab:Parameters}). The mass--temperature diagram in the lower panel of Figure~\ref{fig:dartmouth} indicates that the temperatures of both stars are consistent with predictions, given our uncertainties, and again suggest an age near 2 Gyr.

The results for the Ab go against what has usually been found for other low-mass stars in eclipsing binaries. Most previous studies have shown that M dwarfs in such systems are typically ``inflated'' by up to 10\% or so, and have
their temperatures suppressed by up to about 5$\%$.  This is believed to be due to magnetic activity, as most of those systems are close binaries in which the stars are tidally synchronized and are rotating rapidly. This enhances their activity, which is usually manifested by modulations in the light curve produced by spots, X-ray emission, and other phenomena, and leads to the observed effects \citep[e.g.,][]{2001ApJ...559..353M, 2007A&A...472L..17C}. For the PH1 system  we find, on the other hand, that the size of the Ab is consistent with theoretical expectations, and so is its temperature.  Unfortunately we have no information on the activity level of the Ab, which is some 700 times fainter than the primary.

The current precision of the measurements is about 6$\%$ in the masses and 3--6$\%$ in the radii of the stars in PH1, while the temperatures are only good to about 150\,K, mostly limited by the rotationally broadened spectral lines of the primary. Additional spectroscopic observations and further data from \emph{Kepler} should help to reduce some
of these uncertainties, providing for an even more stringent test of models.

\subsection{Photometric deconvolution}
\label{sec:deconvoultion}

An additional estimate of the contamination level in the $Kp$ band is possible using available multi-color photometry for the system and the stellar evolution models. The photometric measurements in the KIC, as well as the near-infrared magnitudes from 2MASS, contain significant light from four stars: the two components of the eclipsing binary, and those of the 0.7$^{\prime\prime}$ contaminator. We assume here that the 3$^{\prime\prime}$ contaminator is angularly separated enough that these magnitudes are unaffected. With the additional assumption that the four stars are physically associated (see Section \ref{sec:extrastars}), we carried out a deconvolution of all the magnitude measurements using fluxes from a model isochrone, for which we adopted a 2-Gyr Dartmouth model that provides a good fit to the properties of the eclipsing pair (Aa and Ab)(Sect.~\ref{sec:stellarevolution}). Note that those properties (in particular the radii and temperatures of the eclipsing components) depend to some extent on the level of contamination adopted in the photometric-dynamical modeling of the \emph{Kepler} light curve, so we proceeded by iterations between the photometric deconvolution described below, the photometric-dynamical modeling, and the model isochrone fitting.

To deconvolve the combined light of the four stars (SDSS $griz$, 2MASS $JHK_s$, and the custom D51 magnitude from the KIC) we fixed the masses of components Aa and Ab to their  values obtained from the photometric-dynamical  model, which sets their fluxes in all passbands from the isochrone. We then added the masses of Ba and Bb as free parameters, along with the distance modulus and reddening, and performed a grid search on those four variables to find the best match between the predicted and measured values of the combined light. Appropriate reddening corrections were applied to each passband. To constrain the masses of stars Ba and Bb we additionally used our measurement of the brightness difference of their combined light relative to Aa and Ab, from our AO measurements ($\Delta J = 1.89 \pm 0.04$, $\Delta K_s = 1.67 \pm 0.03$). As a further constraint to allow a distinction between Ba and Bb we used a rough estimate of the $K_s$-band brightness difference between those stars from the AO image, of $1.0 \pm 1.0$ mag.

We obtained a good fit to all photometric measurements to well within twice their uncertainties, in most cases, and estimates for the masses of stars Ba and Bb of $\sim$0.99\,$M_{\sun}$ and $\sim$0.51\,$M_{\sun}$, corresponding approximately to spectral types G2 and M2. The fitted distance modulus corresponds to a distance of about 1500 pc.  The combined apparent brightness of Ba and Bb in the \emph{Kepler} band amounts to $10.5 \pm 2.5$\% of the light of Aa and Ab. From its measured brightness of $R = 18.73 \pm 0.05$ the 3$^{\prime\prime}$ companion contributes an additional 1$\%$. This is consistent with our initial estimates of contamination presented in Section \ref{sec:contamination} and assumed in our photometric-dynamical  modeling. Our distance estimate along with the angular separation of the Ba and Bb binary from the target translates to a linear separation of about 1000 AU. Stars Ba and Bb are separated by roughly 60 AU.

\subsection{Stellar variability} 
\label{sec:variability}
In addition to the eclipses and transits, the \emph{Kepler} light curve exhibits low-amplitude variability on a timescale of a few days. With KIC 4862625 providing $\sim$90$\%$ of the flux in the light curve, we attribute this variability to Aa. To further analyze this modulation, we mildly detrended the SAP light curve by omitting the eclipses and removing low-frequency power by using a polynomial or a wide Savitzy-Golay filter (50 cadences on each side). Outliers greater than 4-$\sigma$ from the mean were omitted. A Fourier analysis revealed a strong signal with a period of 2.6397 $\pm$ 0.0014 days; this period is consistent with an autocorrelation analysis. The periodicity is present in all Quarters, and was confirmed by phase folding each Quarter. The amplitude is on average 300 $\pm$ 60 ppm, though its amplitude is noticeably diminished in Quarters 10 and 11. We note that no correction for contamination has been applied in this amplitude estimate. 

In stellar binaries, spin-orbit alignment is expected to be the first to occur. Next is synchronization of the rotation of the two stars with the orbital motion. Orbital circularization is predicted to be the last to occur (see \citeauthor{2008EAS....29....1M} \citeyear{2008EAS....29....1M} and references within). If we assume the modulation is caused by a long-lived cluster of starspots on the primary star (Aa), the periodicity is then the rotation period of the star. Using the radius of the star derived from the photometric-dynamical model and the inclination of the binary, a predicted projected rotation velocity of $v\sin{i}$ = 33.2 $\pm$ 0.9 km~s$^{-1}$ is derived, is consistent with the observed values obtained from the SME and BF analyses (32.6 $\pm$ 2.0 km s$^{-1}$ and  32.38 $\pm$ 0.47 km s$^{-1}$ respectively). This agreement lends credence to the hypothesis that starspot modulation is the source of the periodicity, that the stellar radius is correct, and that the spin axis of the star is mostly aligned with the orbital axis of the binary. It follows that the stellar spin is very far from being tidally locked into a pseudosynchronous spin state; a period of 15.5 days is expected using the formulation of \cite{1981A&A....99..126H}. This implies the binary (Aa and Ab) is a relatively young system with an age less than approximately 9 Gyrs consistent with the findings from the stellar evolution modeling in Section \ref{sec:stellarevolution}.

We note that we cannot rule out a $\gamma$-Doradus type stellar oscillation as the source of the 2.64 day modulation. This alternative hypothesis is attractive because of the sinusoidal shape of the modulation and long-duration stability of the modulation (in amplitude and phase), but the change in amplitude in Quarters 10 and 11 and the excellent agreement with the expected rotation period from $V_{rot} \sin{i}$  make the starspot hypothesis more attractive. The primary eclipse fits do not show correlated residuals indicative of a starspot crossing, so the starspot hypothesis cannot be unambiguously confirmed as was done in the case of Kepler-47 (Orosz et al. 2012b). 

\section{Planet Properties}

Derived physical and orbital parameters for the  transiting planet, PH1b,  are listed in Table \ref{tab:DerivedParameters}. The best-fit radius for PH1b is 6.18 $\pm$ 0.17 R$_{\oplus}$. Without eclipse timing variations on the host stars (Aa and Ab), the photometric-dynamical  model can only place an upper limit on the mass of the planet. At the  99.7$\%$ confidence level, planet's mass must be less than 169 M$_{\oplus}$ (0.531 Jupiter masses). With a sub-Jupiter mass and radius, the transiting body is well within the planetary regime, making PH1b Planet Hunters' first confirmed planet. 

\begin{figure}
\epsscale{1.0}
\plotone{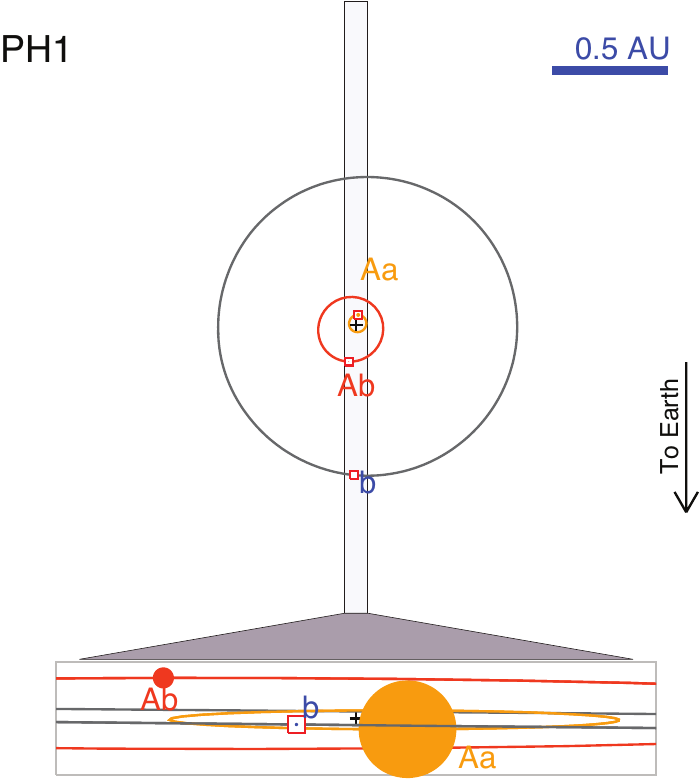}
\caption{Orbital configuration of the PH1 system for the reference epoch, $t_0 = $2,454,970 (BJD).The inner close binary  (primary star Aa and the secondary star Ab) and the planet (labeled b) are shown. The outer visual binary (Ba and Bb) is not depicted.  Top: A scaled, face-on view of the orbits. On this scale the stars and the planet are too small to be seen, and their positions are represented by the boxes. Bottom: The region between the vertical lines in the top diagram is shown on an expanded scale with an orientation corresponding to what would be seen from Earth.}
\label{fig:orbit} 
\end{figure}

Figure \ref{fig:orbit} shows schematic diagrams of the stellar and planetary orbits.  Assuming Aa is the dominant source of insolation, neglecting the small impact of Ab on the surface temperature, we estimate the equilibrium temperature for PH1b. Assuming a Bond albedo of 0.3 (similar to that of Neptune) and emissivity of 0.9, we estimate a minimum and maximum attainable equilibrium temperatures for PH1b. We find  a minimum temperature  of 463 K and a maximum at  498 K. Thus, PH1b is too hot to be in the habitable zone. Although PH1b is a gas giant planet, even if there is a  possibility of rocky moons orbiting the body, their surfaces would be too hot for liquid water to exist. 

The individual planet transits and the best-fitting model are plotted in Figure \ref{fig:photofitlc}. All seven transits of PH1b were across star Aa. With a mutual inclination of 2.619 $\pm$ 0.057$^\circ$, the projected orbits of PH1b and Ab could overlap, but the planet would only transit the face of the Ab near quadratures, making this a rare event. If there is a chance alignment where PH1b does transit across Ab, due to the much fainter magnitude of Ab,  the total transit signal-to-noise of that event would be below the  \emph{Kepler} point-point noise of 203 ppm. Thus transits across Ab would be at too a low signal-to-noise to be identified in the light curve.  

The derived upper mass limit of 169 M$_{\oplus}$ is the mass required to perturb the orbits of the Aa and Ab such that it would induce observable eclipse timing variations, and therefore likely to be much higher than the true mass of the planet. Using the radius-mass relationship derived from \cite{2011ApJS..197....8L} and \cite*[][private communication]{2012PASP..124..323K}, we find a more reasonable mass range for PH1b lies is somewhere around  20-50 M$_{\oplus}$ or 0.08 - 0.14 Jupiter masses.  As noted by \cite{2012ApJ...758...87O}, all  transiting circumbinary planets, including  PH1b, are  Saturn-sized or smaller. This trend may support the  \cite{2008A&A...483..633P} prediction that Jupiter-mass circumbinary planets should be rare due to being  ejected or scattered out to the edge of circumstellar disk, but this result is far from confirmed. The detection biases have yet to be characterized for any of the circumbinary searches, as is the case for Planet Hunters as well. PH1b was found serendipitously, and a systematic search of the Kepler eclipsing binaries has yet to be performed.

\section{Orbital Stability and Formation}

Following the analytic approximation given by \cite{1999AJ....117..621H}, the critical orbital period and radius  in this binary below which the planet's orbit could experience an instability is  respectively 107 days and 0.55 AU. Like the previously discovered \emph{Kepler} circumbinary planets, with a best-fit orbital period of 138.317 days and semimajor axis of 0.652 AU,  the orbit of PH1b is very close to the edge of stability, being only 29\% above the critical period and only 18.6\% larger than the critical orbital radius. Orbital integrations using MERCURY \citep{1999MNRAS.304..793C} of the system using best-fit physical and orbital parameters for Aa, Ab, and PH1b (neglecting the effects of Ba and Bb), confirm that the system is indeed stable on gigayear time scales.

With the long-term stability of the PH1 system, the existence of the planet has implications for formation scenarios of the stellar system and other multistellar systems. Fragmentation and n-body dynamical interactions have been proposed as mechanisms for the formation of stellar systems composed of 3 or more stars. The existence of  PH1b itself suggests that there has been little interaction of the outer binary (Ba and Bb) with the inner eclipsing binary (Aa and Ab). Any dynamical close encounters between (Aa and Ab) and an additional star early on in its formation would have perturbed the surrounding circumbinary disk, likely obliterating the planetesimal disk and preventing accreting planetary cores from forming. Thus Ba and Bb have likely not had a close encounter with Aa and Ab. The existence of PH1b suggests rather than N-body interactions, fragmentation in the molecular cloud is the likely scenario for the formation of the four-star system, preserving the planetesimal disk around Aa and Ab to grow PH1b. 

\section{Conclusions}

In this paper we have characterized the properties of the first confirmed planet in the \emph{Kepler} field found by Planet Hunters citizen science project. PH1b is a new circumbinary planet orbiting an eclipsing binary (KIC 4862625) in a hierarchical quadruple star system. From follow-up observations and analysis of the planetary system we conclude the following:
\begin{enumerate}
\item PH1 is a 6.18 $\pm$ 0.17 R$_{\oplus}$ planet on a 138.317$^{+0.040}_{-0.027}$ day orbit. At the 99.7$\%$ confidence level, we place an upper limit mass of 169 R$_{\oplus}$ (0.531 Jupiter masses).
\item PH1's host stars are an eclipsing binary (Aa and Ab) with a 20.0002468 $\pm$ 0.0000044day period. Aa is a 1.734$\pm$ 0.044 R$_{\odot}$  and  1.528 $\pm$ 0.087 M$_{\odot}$ F dwarf.  Ab is a 0.378 $\pm$ 0.023 R$_{\odot}$  and  0.408 $\pm$ 0.024 M$_{\odot}$ M dwarf.
\item Adaptive optic observations reveal an additional pair of stars in a visual binary (Ba and Bb) contaminating the Kepler photometric aperture, likely composed of a G2 and a M2 stars, at a projected distance of $\sim$1000 AU. Radial velocity observations confirm the pair is associated and bound to the planetary system. 
\item We estimate the age of the planetary system based on stellar evolution models of Aa and Ab at $\sim$2 Gyrs.
\end{enumerate}

The frequency and range of orbital configurations for circumbinary and other multistar planetary systems has yet to be fully explored, but with this seventh confirmed transiting circumbinary planets, we are moving closer to probing this parameter space. These planetary systems provide new boundary conditions for planet formation. This ever increasing sample of dynamically extreme planetary systems will serve as unique and vital tests of proposed planetesimal formation models and core accretion theories  \citep{2000MNRAS.317..773B, 2008A&A...483..633P,
2012ApJ...754L..16P,2012ApJ...752...71M,2012ApJ...757L..29A}. PH1b was found serendipitously by Planet Hunters volunteers examining the light curves of the known \emph{Kepler} eclipsing binaries. The discovery of PH1 highlights the potential of visual inspection through crowd sourcing to identify planet transits in eclipsing multistar systems, where the primary and secondary eclipses of host stars make detection challenging compared to planets with single host stars. In particular, Planet Hunters may exploit a niche identifying circumbinary planetary systems where the planet is not sufficiently massive to produce measurable eclipse timing variations that would be identifiable via automatic detection algorithms. 

\noindent{\it Acknowledgements} 

\noindent The data presented in this paper are the result of the efforts of the Planet Hunters volunteers, without whom this work would not have been possible. Their contributions are individually acknowledged at http://www.planethunters.org/authors. We also acknowledge the following list of individuals who flagged one or more of the  transit events on Talk  discussed in this paper before or after discovery of the planet:  Hans Martin Schwengeler, Dr. Johann Sejpka, and Arvin Joseff Tan.
\\
\\
MES is supported by a National Science Foundation (NSF) Astronomy and Astrophysics Postdoctoral Fellowship under award AST-1003258 and in part by an American Philosophical Society Franklin Grant. JAO and WFW acknowledge support from the \emph{Kepler} Participating Scientist Program via NASA grant NNX12AD23G and the NSF via grant AST-1109928. JAC acknowledges NASA support through Hubble Fellowship grants, awarded by STScI, operated by AURA. DAF acknowledges funding support from Yale University and support from the NASA Supplemental Outreach Award, 10-OUTRCH.210-0001. GT also acknowledges support from NSF through grant AST-1007992. CJL acknowledges support from The Leverhulme Trust. KS acknowledges support from a NASA Einstein Postdoctoral Fellowship grant number PF9-00069, issued by the Chandra X-ray Observatory Center, which is operated by the Smithsonian Astrophysical Observatory for and on behalf of NASA under contract NAS8-03060. The Planet Hunters `Talk' discussion tool used was developed at the Adler Planetarium with support from the NSF CDI grant: DRL-0941610. 
\\
\\
The data presented herein were partly obtained at the W. M. Keck Observatory, which is operated as a scientific partnership among the California Institute of Technology, the University of California, and the National Aeronautics and Space Administration. The Observatory was made possible by the generous financial support of the W. M. Keck Foundation. We thank the observers from the California Planet Survey collaboration for their efforts. We thank Mike Brown and Emily Schaller for their assistance in obtaining NGS AO observations. We thank Tom Barclay, Gibor Basri, Luke Dones, Hans Kjeldsen, Hal Levison, and Andrej Pr\^sa for useful discussions. 
\\
\\
This paper includes data collected by the \emph{Kepler} spacecraft, and we gratefully acknowledge the entire \emph{Kepler} mission team's efforts in obtaining and providing the light curves used in this analysis. Funding for the \emph{Kepler} mission is provided by the NASA Science Mission directorate. The publicly released \emph{Kepler} light curves were obtained from the Mikulski Archive for Space Telescopes (MAST). STScI is operated by the Association of Universities for Research in Astronomy, Inc., under NASA contract NAS5-26555. Support for MAST for non-HST data is provided by the NASA Office of Space Science via grant NNX09AF08G and by other grants and contracts.

\noindent {\it Facilities:} \facility{Kepler} \facility{Keck:I}  \facility{Keck:II} \facility{SARA}

\bibliographystyle{apj}

\begin{thebibliography}{74}
\expandafter\ifx\csname natexlab\endcsname\relax\def\natexlab#1{#1}\fi

\bibitem[{{Adams} {et~al.}(2012){Adams}, {Ciardi}, {Dupree}, {Gautier},
  {Kulesa}, \& {McCarthy}}]{2012AJ....144...42A}
{Adams}, E.~R., {Ciardi}, D.~R., {Dupree}, A.~K., {et~al.} 2012, \aj, 144, 42

\bibitem[{{Alexander}(2012)}]{2012ApJ...757L..29A}
{Alexander}, R. 2012, \apjl, 757, L29

\bibitem[{{Andersen}(1991)}]{1991A&ARv...3...91A}
{Andersen}, J. 1991, \aapr, 3, 91

\bibitem[{{Bass} {et~al.}(2012){Bass}, {Orosz}, {Welsh}, {Windmiller}, {Ames
  Gregg}, {Fetherolf}, {Wade}, \& {Quinn}}]{2012ApJ...761..157B}
{Bass}, G., {Orosz}, J.~A., {Welsh}, W.~F., {et~al.} 2012, \apj, 761, 157

\bibitem[{{Batalha} {et~al.}(2013){Batalha}, {Rowe}, {Bryson}, {Barclay},
  {Burke}, {Caldwell}, {Christiansen}, {Mullally}, {Thompson}, {Brown},
  {Dupree}, {Fabrycky}, {Ford}, {Fortney}, {Gilliland}, {Isaacson}, {Latham},
  {Marcy}, {Quinn}, {Ragozzine}, {Shporer}, {Borucki}, {Ciardi}, {Gautier},
  {Haas}, {Jenkins}, {Koch}, {Lissauer}, {Rapin}, {Basri}, {Boss}, {Buchhave},
  {Carter}, {Charbonneau}, {Christensen-Dalsgaard}, {Clarke}, {Cochran},
  {Demory}, {Desert}, {Devore}, {Doyle}, {Esquerdo}, {Everett}, {Fressin},
  {Geary}, {Girouard}, {Gould}, {Hall}, {Holman}, {Howard}, {Howell},
  {Ibrahim}, {Kinemuchi}, {Kjeldsen}, {Klaus}, {Li}, {Lucas}, {Meibom},
  {Morris}, {Pr{\v s}a}, {Quintana}, {Sanderfer}, {Sasselov}, {Seader},
  {Smith}, {Steffen}, {Still}, {Stumpe}, {Tarter}, {Tenenbaum}, {Torres},
  {Twicken}, {Uddin}, {Van Cleve}, {Walkowicz}, \&
  {Welsh}}]{2012arXiv1202.5852B}
{Batalha}, N.~M., {Rowe}, J.~F., {Bryson}, S.~T., {et~al.} 2013, \apjs, 204, 24

\bibitem[{{Bate} {et~al.}(2000){Bate}, {Bonnell}, {Clarke}, {Lubow}, {Ogilvie},
  {Pringle}, \& {Tout}}]{2000MNRAS.317..773B}
{Bate}, M.~R., {Bonnell}, I.~A., {Clarke}, C.~J., {et~al.} 2000, \mnras, 317,
  773

\bibitem[{{Bennett} {et~al.}(1999){Bennett}, {Rhie}, {Becker}, {Butler},
  {Dann}, {Kaspi}, {Leibowitz}, {Lipkin}, {Maoz}, {Mendelson}, {Peterson},
  {Quinn}, {Shemmer}, {Thomson}, \& {Turner}}]{1999Natur.402...57B}
{Bennett}, D.~P., {Rhie}, S.~H., {Becker}, A.~C., {et~al.} 1999, \nat, 402, 57

\bibitem[{{Beuermann} {et~al.}(2012){Beuermann}, {Dreizler}, {Hessman}, \&
  {Deller}}]{2012A&A...543A.138B}
{Beuermann}, K., {Dreizler}, S., {Hessman}, F.~V., \& {Deller}, J. 2012, \aap,
  543, A138

\bibitem[{{Borucki} {et~al.}(2010){Borucki}, {Koch}, {Basri}, {Batalha},
  {Brown}, {Caldwell}, {Caldwell}, {Christensen-Dalsgaard}, {Cochran},
  {DeVore}, {Dunham}, {Dupree}, {Gautier}, {Geary}, {Gilliland}, {Gould},
  {Howell}, {Jenkins}, {Kondo}, {Latham}, {Marcy}, {Meibom}, {Kjeldsen},
  {Lissauer}, {Monet}, {Morrison}, {Sasselov}, {Tarter}, {Boss}, {Brownlee},
  {Owen}, {Buzasi}, {Charbonneau}, {Doyle}, {Fortney}, {Ford}, {Holman},
  {Seager}, {Steffen}, {Welsh}, {Rowe}, {Anderson}, {Buchhave}, {Ciardi},
  {Walkowicz}, {Sherry}, {Horch}, {Isaacson}, {Everett}, {Fischer}, {Torres},
  {Johnson}, {Endl}, {MacQueen}, {Bryson}, {Dotson}, {Haas}, {Kolodziejczak},
  {Van Cleve}, {Chandrasekaran}, {Twicken}, {Quintana}, {Clarke}, {Allen},
  {Li}, {Wu}, {Tenenbaum}, {Verner}, {Bruhweiler}, {Barnes}, \&
  {Prsa}}]{2010Sci...327..977B}
{Borucki}, W.~J., {Koch}, D., {Basri}, G., {et~al.} 2010, Science, 327, 977

\bibitem[{{Brown} {et~al.}(2011){Brown}, {Latham}, {Everett}, \&
  {Esquerdo}}]{2011AJ....142..112B}
{Brown}, T.~M., {Latham}, D.~W., {Everett}, M.~E., \& {Esquerdo}, G.~A. 2011,
  \aj, 142, 112

\bibitem[{{Chabrier} {et~al.}(2007){Chabrier}, {Gallardo}, \&
  {Baraffe}}]{2007A&A...472L..17C}
{Chabrier}, G., {Gallardo}, J., \& {Baraffe}, I. 2007, \aap, 472, L17

\bibitem[{{Chambers}(1999)}]{1999MNRAS.304..793C}
{Chambers}, J.~E. 1999, \mnras, 304, 793

\bibitem[{{Christiansen} {et~al.}(2011){Christiansen}, {Van Cleve}, {Jenkins},
  {Caldwell}, {Allen}, \& {Barclay}}]{Keplerdatamanual}
{Christiansen}, J.~L., {Van Cleve}, J.~E., {Jenkins}, J.~M., {et~al.} 2011, {
  Kepler Data Characteristics Handbook (KSCI-19040-002)}

\bibitem[{{Chubak} {et~al.}(2012){Chubak}, {Marcy}, {Fischer}, {Howard},
  {Isaacson}, {Johnson}, \& {Wright}}]{2012arXiv1207.6212C}
{Chubak}, C., {Marcy}, G., {Fischer}, D.~A., {et~al.} 2012, ArXiv e-prints

\bibitem[{{Correia} {et~al.}(2006){Correia}, {Zinnecker}, {Ratzka}, \&
  {Sterzik}}]{2006A&A...459..909C}
{Correia}, S., {Zinnecker}, H., {Ratzka}, T., \& {Sterzik}, M.~F. 2006, \aap,
  459, 909

\bibitem[{{Deeg} {et~al.}(2008){Deeg}, {Oca{\~n}a}, {Kozhevnikov},
  {Charbonneau}, {O'Donovan}, \& {Doyle}}]{2008A&A...480..563D}
{Deeg}, H.~J., {Oca{\~n}a}, B., {Kozhevnikov}, V.~P., {et~al.} 2008, \aap, 480,
  563

\bibitem[{{Deeg} {et~al.}(1998){Deeg}, {Doyle}, {Kozhevnikov}, {Martin},
  {Oetiker}, {Palaiologou}, {Schneider}, {Afonso}, {Dunham}, {Jenkins},
  {Ninkov}, {Stone}, \& {Zakharova}}]{1998A&A...338..479D}
{Deeg}, H.~J., {Doyle}, L.~R., {Kozhevnikov}, V.~P., {et~al.} 1998, \aap, 338,
  479

\bibitem[{{Dotter} {et~al.}(2008){Dotter}, {Chaboyer}, {Jevremovi{\'c}},
  {Kostov}, {Baron}, \& {Ferguson}}]{2008ApJS..178...89D}
{Dotter}, A., {Chaboyer}, B., {Jevremovi{\'c}}, D., {et~al.} 2008, \apjs, 178,
  89

\bibitem[{{Doyle} {et~al.}(2000){Doyle}, {Deeg}, {Kozhevnikov}, {Oetiker},
  {Mart{\'{\i}}n}, {Blue}, {Rottler}, {Stone}, {Ninkov}, {Jenkins},
  {Schneider}, {Dunham}, {Doyle}, \& {Paleologou}}]{2000ApJ...535..338D}
{Doyle}, L.~R., {Deeg}, H.~J., {Kozhevnikov}, V.~P., {et~al.} 2000, \apj, 535,
  338

\bibitem[{{Doyle} {et~al.}(2011){Doyle}, {Carter}, {Fabrycky}, {Slawson},
  {Howell}, {Winn}, {Orosz}, {Pr{\v s}a}, {Welsh}, {Quinn}, {Latham}, {Torres},
  {Buchhave}, {Marcy}, {Fortney}, {Shporer}, {Ford}, {Lissauer}, {Ragozzine},
  {Rucker}, {Batalha}, {Jenkins}, {Borucki}, {Koch}, {Middour}, {Hall},
  {McCauliff}, {Fanelli}, {Quintana}, {Holman}, {Caldwell}, {Still},
  {Stefanik}, {Brown}, {Esquerdo}, {Tang}, {Furesz}, {Geary}, {Berlind},
  {Calkins}, {Short}, {Steffen}, {Sasselov}, {Dunham}, {Cochran}, {Boss},
  {Haas}, {Buzasi}, \& {Fischer}}]{2011Sci...333.1602D}
{Doyle}, L.~R., {Carter}, J.~A., {Fabrycky}, D.~C., {et~al.} 2011, Science,
  333, 1602

\bibitem[{{Fischer} {et~al.}(2012){Fischer}, {Schwamb}, {Schawinski},
  {Lintott}, {Brewer}, {Giguere}, {Lynn}, {Parrish}, {Sartori}, {Simpson},
  {Smith}, {Spronck}, {Batalha}, {Rowe}, {Jenkins}, {Bryson}, {Prsa},
  {Tenenbaum}, {Crepp}, {Morton}, {Howard}, {Beleu}, {Kaplan}, {Vannispen},
  {Sharzer}, {Defouw}, {Hajduk}, {Neal}, {Nemec}, {Schuepbach}, \&
  {Zimmermann}}]{2012MNRAS.419.2900F}
{Fischer}, D.~A., {Schwamb}, M.~E., {Schawinski}, K., {et~al.} 2012, \mnras,
  419, 2900

\bibitem[{{Ghez} {et~al.}(2008){Ghez}, {Salim}, {Weinberg}, {Lu}, {Do}, {Dunn},
  {Matthews}, {Morris}, {Yelda}, {Becklin}, {Kremenek}, {Milosavljevic}, \&
  {Naiman}}]{2008ApJ...689.1044G}
{Ghez}, A.~M., {Salim}, S., {Weinberg}, N.~N., {et~al.} 2008, \apj, 689, 1044

\bibitem[{{Gim{\'e}nez}(2006)}]{2006A&A...450.1231G}
{Gim{\'e}nez}, A. 2006, \aap, 450, 1231

\bibitem[{{Heisler} \& {Tremaine}(1986)}]{1986Icar...65...13H}
{Heisler}, J., \& {Tremaine}, S. 1986, \icarus, 65, 13

\bibitem[{{Holman} \& {Wiegert}(1999)}]{1999AJ....117..621H}
{Holman}, M.~J., \& {Wiegert}, P.~A. 1999, \aj, 117, 621

\bibitem[{{Howard} {et~al.}(2010){Howard}, {Johnson}, {Marcy}, {Fischer},
  {Wright}, {Bernat}, {Henry}, {Peek}, {Isaacson}, {Apps}, {Endl}, {Cochran},
  {Valenti}, {Anderson}, \& {Piskunov}}]{2010ApJ...721.1467H}
{Howard}, A.~W., {Johnson}, J.~A., {Marcy}, G.~W., {et~al.} 2010, \apj, 721,
  1467

\bibitem[{{Howell} {et~al.}(2012){Howell}, {Rowe}, {Bryson}, {Quinn}, {Marcy},
  {Isaacson}, {Ciardi}, {Chaplin}, {Metcalfe}, {Monteiro}, {Appourchaux},
  {Basu}, {Creevey}, {Gilliland}, {Quirion}, {Stello}, {Kjeldsen},
  {Christensen-Dalsgaard}, {Elsworth}, {Garc{\'{\i}}a}, {Houdek}, {Karoff},
  {Molenda-{\.Z}akowicz}, {Thompson}, {Verner}, {Torres}, {Fressin}, {Crepp},
  {Adams}, {Dupree}, {Sasselov}, {Dressing}, {Borucki}, {Koch}, {Lissauer},
  {Latham}, {Buchhave}, {Gautier}, {Everett}, {Horch}, {Batalha}, {Dunham},
  {Szkody}, {Silva}, {Mighell}, {Holberg}, {Ballot}, {Bedding}, {Bruntt},
  {Campante}, {Handberg}, {Hekker}, {Huber}, {Mathur}, {Mosser}, {R{\'e}gulo},
  {White}, {Christiansen}, {Middour}, {Haas}, {Hall}, {Jenkins}, {McCaulif},
  {Fanelli}, {Kulesa}, {McCarthy}, \& {Henze}}]{2012ApJ...746..123H}
{Howell}, S.~B., {Rowe}, J.~F., {Bryson}, S.~T., {et~al.} 2012, \apj, 746, 123

\bibitem[{{Hut}(1981)}]{1981A&A....99..126H}
{Hut}, P. 1981, \aap, 99, 126

\bibitem[{{Jenkins} {et~al.}(2002){Jenkins}, {Caldwell}, \&
  {Borucki}}]{2002ApJ...564..495J}
{Jenkins}, J.~M., {Caldwell}, D.~A., \& {Borucki}, W.~J. 2002, \apj, 564, 495

\bibitem[{{Jenkins} {et~al.}(2010{\natexlab{a}}){Jenkins}, {Caldwell},
  {Chandrasekaran}, {Twicken}, {Bryson}, {Quintana}, {Clarke}, {Li}, {Allen},
  {Tenenbaum}, {Wu}, {Klaus}, {Van Cleve}, {Dotson}, {Haas}, {Gilliland},
  {Koch}, \& {Borucki}}]{2010ApJ...713L.120J}
{Jenkins}, J.~M., {Caldwell}, D.~A., {Chandrasekaran}, H., {et~al.}
  2010{\natexlab{a}}, \apjl, 713, L120

\bibitem[{{Jenkins} {et~al.}(2010{\natexlab{b}}){Jenkins}, {Caldwell},
  {Chandrasekaran}, {Twicken}, {Bryson}, {Quintana}, {Clarke}, {Li}, {Allen},
  {Tenenbaum}, {Wu}, {Klaus}, {Middour}, {Cote}, {McCauliff}, {Girouard},
  {Gunter}, {Wohler}, {Sommers}, {Hall}, {Uddin}, {Wu}, {Bhavsar}, {Van Cleve},
  {Pletcher}, {Dotson}, {Haas}, {Gilliland}, {Koch}, \&
  {Borucki}}]{2010ApJ...713L..87J}
---. 2010{\natexlab{b}}, \apjl, 713, L87

\bibitem[{{Jiang} \& {Tremaine}(2010)}]{2010MNRAS.401..977J}
{Jiang}, Y.-F., \& {Tremaine}, S. 2010, \mnras, 401, 977

\bibitem[{{Kaib} {et~al.}(2013){Kaib}, {Raymond}, \&
  {Duncan}}]{2013Natur.493..381K}
{Kaib}, N.~A., {Raymond}, S.~N., \& {Duncan}, M. 2013, \nat, 493, 381

\bibitem[{{Kane} \& {Gelino}(2012)}]{2012PASP..124..323K}
{Kane}, S.~R., \& {Gelino}, D.~M. 2012, \pasp, 124, 323

\bibitem[{{Kostov} {et~al.}(2012){Kostov}, {McCullough}, {Hinse}, {Tsvetanov},
  {H{\'e}brard}, {D{\'{\i}}az}, {Deleuil}, \& {Valenti}}]{2012arXiv1210.3850K}
{Kostov}, V.~B., {McCullough}, P., {Hinse}, T., {et~al.} 2012, ArXiv e-prints

\bibitem[{{Kouwenhoven} {et~al.}(2010){Kouwenhoven}, {Goodwin}, {Parker},
  {Davies}, {Malmberg}, \& {Kroupa}}]{2010MNRAS.404.1835K}
{Kouwenhoven}, M.~B.~N., {Goodwin}, S.~P., {Parker}, R.~J., {et~al.} 2010,
  \mnras, 404, 1835

\bibitem[{{Lee} {et~al.}(2009){Lee}, {Kim}, {Kim}, {Koch}, {Lee}, {Kim}, \&
  {Park}}]{2009AJ....137.3181L}
{Lee}, J.~W., {Kim}, S.-L., {Kim}, C.-H., {et~al.} 2009, \aj, 137, 3181

\bibitem[{{Lillo-Box} {et~al.}(2012){Lillo-Box}, {Barrado}, \&
  {Bouy}}]{2012A&A...546A..10L}
{Lillo-Box}, J., {Barrado}, D., \& {Bouy}, H. 2012, \aap, 546, A10

\bibitem[{{Lintott} {et~al.}(2012){Lintott}, {Schwamb}, {Sharzer}, {Fischer},
  {Barclay}, {Parrish}, {Batalha}, {Bryson}, {Jenkins}, {Ragozzine}, {Rowe},
  {Schawinski}, {Gagliano}, {Gilardi}, {Jek}, {P{\"a}{\"a}kk{\"o}nen}, \&
  {Smits}}]{2012arXiv1202.6007L}
{Lintott}, C., {Schwamb}, M.~E., {Sharzer}, C., {et~al.} 2012, arXiv:1202.6007

\bibitem[{{Lintott} {et~al.}(2008){Lintott}, {Schawinski}, {Slosar}, {Land},
  {Bamford}, {Thomas}, {Raddick}, {Nichol}, {Szalay}, {Andreescu}, {Murray}, \&
  {Vandenberg}}]{2008MNRAS.389.1179L}
{Lintott}, C.~J., {Schawinski}, K., {Slosar}, A., {et~al.} 2008, \mnras, 389,
  1179

\bibitem[{{Lissauer} {et~al.}(2011){Lissauer}, {Ragozzine}, {Fabrycky},
  {Steffen}, {Ford}, {Jenkins}, {Shporer}, {Holman}, {Rowe}, {Quintana},
  {Batalha}, {Borucki}, {Bryson}, {Caldwell}, {Carter}, {Ciardi}, {Dunham},
  {Fortney}, {Gautier}, {Howell}, {Koch}, {Latham}, {Marcy}, {Morehead}, \&
  {Sasselov}}]{2011ApJS..197....8L}
{Lissauer}, J.~J., {Ragozzine}, D., {Fabrycky}, D.~C., {et~al.} 2011, \apjs,
  197, 8

\bibitem[{{L{\'o}pez-Morales} \& {Ribas}(2005)}]{2005ApJ...631.1120L}
{L{\'o}pez-Morales}, M., \& {Ribas}, I. 2005, \apj, 631, 1120

\bibitem[{{Mazeh}(2008)}]{2008EAS....29....1M}
{Mazeh}, T. 2008, in EAS Publications Series, Vol.~29, EAS Publications Series,
  ed. M.-J. {Goupil} \& J.-P. {Zahn}, 1--65

\bibitem[{{Meschiari}(2012{\natexlab{a}})}]{2012ApJ...752...71M}
{Meschiari}, S. 2012{\natexlab{a}}, \apj, 752, 71

\bibitem[{{Meschiari}(2012{\natexlab{b}})}]{2012ApJ...761L...7M}
---. 2012{\natexlab{b}}, \apjl, 761, L7

\bibitem[{{Morales} {et~al.}(2009){Morales}, {Ribas}, {Jordi}, {Torres},
  {Gallardo}, {Guinan}, {Charbonneau}, {Wolf}, {Latham}, {Anglada-Escud{\'e}},
  {Bradstreet}, {Everett}, {O'Donovan}, {Mandushev}, \&
  {Mathieu}}]{2009ApJ...691.1400M}
{Morales}, J.~C., {Ribas}, I., {Jordi}, C., {et~al.} 2009, \apj, 691, 1400

\bibitem[{{Mullan} \& {MacDonald}(2001)}]{2001ApJ...559..353M}
{Mullan}, D.~J., \& {MacDonald}, J. 2001, \apj, 559, 353

\bibitem[{{Orosz} \& {Hauschildt}(2000)}]{2000A&A...364..265O}
{Orosz}, J.~A., \& {Hauschildt}, P.~H. 2000, \aap, 364, 265

\bibitem[{{Orosz} {et~al.}(2012{\natexlab{a}}){Orosz}, {Welsh}, {Carter},
  {Fabrycky}, {Cochran}, {Endl}, {Ford}, {Haghighipour}, {MacQueen}, {Mazeh},
  {Sanchis-Ojeda}, {Short}, {Torres}, {Agol}, {Buchhave}, {Doyle}, {Isaacson},
  {Lissauer}, {Marcy}, {Shporer}, {Windmiller}, {Barclay}, {Boss}, {Clarke},
  {Fortney}, {Geary}, {Holman}, {Huber}, {Jenkins}, {Kinemuchi}, {Kruse},
  {Ragozzine}, {Sasselov}, {Still}, {Tenenbaum}, {Uddin}, {Winn}, {Koch}, \&
  {Borucki}}]{2012Sci...337.1511O}
{Orosz}, J.~A., {Welsh}, W.~F., {Carter}, J.~A., {et~al.} 2012{\natexlab{a}},
  Science, 337, 1511

\bibitem[{{Orosz} {et~al.}(2012{\natexlab{b}}){Orosz}, {Welsh}, {Carter},
  {Brugamyer}, {Buchhave}, {Cochran}, {Endl}, {Ford}, {MacQueen}, {Short},
  {Torres}, {Windmiller}, {Agol}, {Barclay}, {Caldwell}, {Clarke}, {Doyle},
  {Fabrycky}, {Geary}, {Haghighipour}, {Holman}, {Ibrahim}, {Jenkins},
  {Kinemuchi}, {Li}, {Lissauer}, {Pr{\v s}a}, {Ragozzine}, {Shporer}, {Still},
  \& {Wade}}]{2012ApJ...758...87O}
---. 2012{\natexlab{b}}, \apj, 758, 87

\bibitem[{{Paardekooper} {et~al.}(2012){Paardekooper}, {Leinhardt},
  {Th{\'e}bault}, \& {Baruteau}}]{2012ApJ...754L..16P}
{Paardekooper}, S.-J., {Leinhardt}, Z.~M., {Th{\'e}bault}, P., \& {Baruteau},
  C. 2012, \apjl, 754, L16

\bibitem[{{Pierens} \& {Nelson}(2008)}]{2008A&A...483..633P}
{Pierens}, A., \& {Nelson}, R.~P. 2008, \aap, 483, 633

\bibitem[{{Pribulla} {et~al.}(2009){Pribulla}, {Rucinski}, {DeBond}, {De
  Ridder}, {Karmo}, {Thomson}, {Croll}, {Og{\l}oza}, {Pilecki}, \&
  {Siwak}}]{2009AJ....137.3646P}
{Pribulla}, T., {Rucinski}, S.~M., {DeBond}, H., {et~al.} 2009, \aj, 137, 3646

\bibitem[{{Pr{\v s}a} {et~al.}(2011){Pr{\v s}a}, {Batalha}, {Slawson}, {Doyle},
  {Welsh}, {Orosz}, {Seager}, {Rucker}, {Mjaseth}, {Engle}, {Conroy},
  {Jenkins}, {Caldwell}, {Koch}, \& {Borucki}}]{2011AJ....141...83P}
{Pr{\v s}a}, A., {Batalha}, N., {Slawson}, R.~W., {et~al.} 2011, \aj, 141, 83

\bibitem[{{Ribas}(2003)}]{2003A&A...398..239R}
{Ribas}, I. 2003, \aap, 398, 239

\bibitem[{{Rucinski}(1992)}]{1992AJ....104.1968R}
{Rucinski}, S.~M. 1992, \aj, 104, 1968

\bibitem[{{Schwamb} {et~al.}(2012){Schwamb}, {Lintott}, {Fischer}, {Giguere},
  {Lynn}, {Smith}, {Brewer}, {Parrish}, {Schawinski}, \&
  {Simpson}}]{2012ApJ...754..129S}
{Schwamb}, M.~E., {Lintott}, C.~J., {Fischer}, D.~A., {et~al.} 2012, \apj, 754,
  129

\bibitem[{{Skrutskie} {et~al.}(2006){Skrutskie}, {Cutri}, {Stiening},
  {Weinberg}, {Schneider}, {Carpenter}, {Beichman}, {Capps}, {Chester},
  {Elias}, {Huchra}, {Liebert}, {Lonsdale}, {Monet}, {Price}, {Seitzer},
  {Jarrett}, {Kirkpatrick}, {Gizis}, {Howard}, {Evans}, {Fowler}, {Fullmer},
  {Hurt}, {Light}, {Kopan}, {Marsh}, {McCallon}, {Tam}, {Van Dyk}, \&
  {Wheelock}}]{2006AJ....131.1163S}
{Skrutskie}, M.~F., {Cutri}, R.~M., {Stiening}, R., {et~al.} 2006, \aj, 131,
  1163

\bibitem[{{Slawson} {et~al.}(2011){Slawson}, {Pr{\v s}a}, {Welsh}, {Orosz},
  {Rucker}, {Batalha}, {Doyle}, {Engle}, {Conroy}, {Coughlin}, {Gregg},
  {Fetherolf}, {Short}, {Windmiller}, {Fabrycky}, {Howell}, {Jenkins}, {Uddin},
  {Mullally}, {Seader}, {Thompson}, {Sanderfer}, {Borucki}, \&
  {Koch}}]{2011AJ....142..160S}
{Slawson}, R.~W., {Pr{\v s}a}, A., {Welsh}, W.~F., {et~al.} 2011, \aj, 142, 160

\bibitem[{{Smith} {et~al.}(2011){Smith}, {Lynn}, {Sullivan}, {Lintott},
  {Nugent}, {Botyanszki}, {Kasliwal}, {Quimby}, {Bamford}, {Fortson},
  {Schawinski}, {Hook}, {Blake}, {Podsiadlowski}, {J{\"o}nsson}, {Gal-Yam},
  {Arcavi}, {Howell}, {Bloom}, {Jacobsen}, {Kulkarni}, {Law}, {Ofek}, \&
  {Walters}}]{2011MNRAS.412.1309S}
{Smith}, A.~M., {Lynn}, S., {Sullivan}, M., {et~al.} 2011, \mnras, 412, 1309

\bibitem[{{Smith} {et~al.}(2012){Smith}, {Stumpe}, {Van Cleve}, {Jenkins},
  {Barclay}, {Fanelli}, {Girouard}, {Kolodziejczak}, {McCauliff}, {Morris}, \&
  {Twicken}}]{2012PASP..124.1000S}
{Smith}, J.~C., {Stumpe}, M.~C., {Van Cleve}, J.~E., {et~al.} 2012, \pasp, 124,
  1000

\bibitem[{{Steffen} {et~al.}(2011){Steffen}, {Quinn}, {Borucki}, {Brugamyer},
  {Bryson}, {Buchhave}, {Cochran}, {Endl}, {Fabrycky}, {Ford}, {Holman},
  {Jenkins}, {Koch}, {Latham}, {MacQueen}, {Mullally}, {Pr{\v s}a},
  {Ragozzine}, {Rowe}, {Sanderfer}, {Seader}, {Short}, {Shporer}, {Thompson},
  {Torres}, {Twicken}, {Welsh}, \& {Windmiller}}]{2011MNRAS.417L..31S}
{Steffen}, J.~H., {Quinn}, S.~N., {Borucki}, W.~J., {et~al.} 2011, \mnras, 417,
  L31

\bibitem[{{Tenenbaum} {et~al.}(2012){Tenenbaum}, {Christiansen}, {Jenkins},
  {Rowe}, {Seader}, {Caldwell}, {Clarke}, {Li}, {Quintana}, {Smith}, {Stumpe},
  {Thompson}, {Twicken}, {Van Cleve}, {Borucki}, {Cote}, {Haas}, {Sanderfer},
  {Girouard}, {Klaus}, {Middour}, {Wohler}, {Batalha}, {Barclay}, \&
  {Nickerson}}]{2012arXiv1201.1048T}
{Tenenbaum}, P., {Christiansen}, J.~L., {Jenkins}, J.~M., {et~al.} 2012, \apjs,
  199, 24

\bibitem[{{Tokovinin}(2001)}]{2001IAUS..200...84T}
{Tokovinin}, A. 2001, in IAU Symposium, Vol. 200, The Formation of Binary
  Stars, ed. H.~{Zinnecker} \& R.~{Mathieu}, 84--92

\bibitem[{{Tokovinin}(1997)}]{1997A&AS..124...75T}
{Tokovinin}, A.~A. 1997, \aaps, 124, 75

\bibitem[{{Torres}(2012)}]{2012arXiv1209.1279T}
{Torres}, G. 2012, ArXiv e-prints

\bibitem[{{Torres} {et~al.}(2010){Torres}, {Andersen}, \&
  {Gim{\'e}nez}}]{2010A&ARv..18...67T}
{Torres}, G., {Andersen}, J., \& {Gim{\'e}nez}, A. 2010, \aapr, 18, 67

\bibitem[{{Torres} \& {Ribas}(2002)}]{2002ApJ...567.1140T}
{Torres}, G., \& {Ribas}, I. 2002, \apj, 567, 1140

\bibitem[{{Valenti} \& {Fischer}(2005)}]{2005ApJS..159..141V}
{Valenti}, J.~A., \& {Fischer}, D.~A. 2005, \apjs, 159, 141

\bibitem[{{Valenti} \& {Piskunov}(1996)}]{1996A&AS..118..595V}
{Valenti}, J.~A., \& {Piskunov}, N. 1996, \aaps, 118, 595

\bibitem[{{Vogt} {et~al.}(1994){Vogt}, {Allen}, {Bigelow}, {Bresee}, {Brown},
  {Cantrall}, {Conrad}, {Couture}, {Delaney}, {Epps}, {Hilyard}, {Hilyard},
  {Horn}, {Jern}, {Kanto}, {Keane}, {Kibrick}, {Lewis}, {Osborne},
  {Pardeilhan}, {Pfister}, {Ricketts}, {Robinson}, {Stover}, {Tucker}, {Ward},
  \& {Wei}}]{1994SPIE.2198..362V}
{Vogt}, S.~S., {Allen}, S.~L., {Bigelow}, B.~C., {et~al.} 1994, in Society of
  Photo-Optical Instrumentation Engineers (SPIE) Conference Series, Vol. 2198,
  Society of Photo-Optical Instrumentation Engineers (SPIE) Conference Series,
  ed. D.~L. {Crawford} \& E.~R. {Craine}, 362

\bibitem[{{Welsh} {et~al.}(2012){Welsh}, {Orosz}, {Carter}, {Fabrycky}, {Ford},
  {Lissauer}, {Pr{\v s}a}, {Quinn}, {Ragozzine}, {Short}, {Torres}, {Winn},
  {Doyle}, {Barclay}, {Batalha}, {Bloemen}, {Brugamyer}, {Buchhave},
  {Caldwell}, {Caldwell}, {Christiansen}, {Ciardi}, {Cochran}, {Endl},
  {Fortney}, {Gautier}, {Gilliland}, {Haas}, {Hall}, {Holman}, {Howard},
  {Howell}, {Isaacson}, {Jenkins}, {Klaus}, {Latham}, {Li}, {Marcy}, {Mazeh},
  {Quintana}, {Robertson}, {Shporer}, {Steffen}, {Windmiller}, {Koch}, \&
  {Borucki}}]{2012Natur.481..475W}
{Welsh}, W.~F., {Orosz}, J.~A., {Carter}, J.~A., {et~al.} 2012, \nat, 481, 475

\bibitem[{{Wielen}(1977)}]{1977A&A....60..263W}
{Wielen}, R. 1977, \aap, 60, 263

\bibitem[{{Yi} {et~al.}(2001){Yi}, {Demarque}, {Kim}, {Lee}, {Ree}, {Lejeune},
  \& {Barnes}}]{2001ApJS..136..417Y}
{Yi}, S., {Demarque}, P., {Kim}, Y.-C., {et~al.} 2001, \apjs, 136, 417

\end{thebibliography}

\clearpage 

\begin{table}[h!]
\centering
\caption{Reduced Kepler light curve for KIC 4862625 for Quarters 1-11 used to measure stellar eclipse timing variations, timing offsets of the planet transits, and changes in planet transit duration described in Section \ref{sec:lightcurve}. This table s published in its entirety in the online edition. A portion is shown here for guidance regarding its form and content.}
\begin{tabular}{l l l }
\hline
\hline
(BJD- 2455000) & Relative Flux & Error \\
 \hline 
         
-35.4887090 & 0.99985  & 0.00017 \\
-35.4682743 & 1.00001  & 0.00017 \\
-35.4478397 & 1.00023  & 0.00017 \\
-35.4274052 & 1.00007  & 0.00017 \\
-35.4069706 & 0.99982  & 0.00017 \\
-35.3865359 & 0.99992  & 0.00017 \\
-35.3661014 & 1.00012  & 0.00017 \\
-35.3456668 & 1.00003  & 0.00017 \\
-35.3252321 & 0.99990  & 0.00017 \\
-35.3047974 & 1.00001  & 0.00017 \\
-35.2843629 & 1.00017  & 0.00017 \\
-35.2639283 & 0.99989  & 0.00017 \\
-35.2434936 & 0.99954  & 0.00017 \\
-35.2230591 & 1.00031  & 0.00017 \\
-35.2026245 & 0.99997  & 0.00017 \\
-35.1821898 & 0.99991  & 0.00017 \\
-35.1617553 & 1.00007  & 0.00017 \\
-35.1413207 & 0.99981  & 0.00017 \\
-35.1208860 & 1.00000  & 0.00017 \\

\hline
\hline

\end{tabular}

\label{tab:lightcurve}
\end{table}

\begin{table}[h!]
\centering
\caption{Measured transit times for PH1b} 
\begin{tabular}{c  c c }
\hline
\hline
Quarter &  Transit Midpoint & 1-$\sigma$ Uncertainty \\
& (BJD- 2455000) & (Minutes) \\
 \hline 
         
2 & 70.80674   &     8.94  \\
4 & 207.44916  &     9.00 \\
 5 & 344.11218  &     6.14  \\
7 & 479.98746   &   8.89\\
8 &  613.17869    &     9.04\\
10& 749.20260    &    6.49 \\
11 &   885.91042   &   8.94 \\
\hline
\hline

\end{tabular}

\label{tab:transittimes}

\end{table}

\begin{longtable}{crrcrr}
 \caption[short]{ Measured  primary and  secondary eclipse times for KIC 4862625 } \\ 
\hline
\hline
Orbital& \multicolumn{2}{c}{Primary Eclipse} & & \multicolumn{2}{c}{Secondary Eclipse } \\
\cline{2-3} \cline{5-6}
   Cycle & \multicolumn{1}{c}{Midpoint} & Uncertainty  & &  \multicolumn{1}{c}{Midpoint}& \multicolumn{1}{c}{Uncertainty} \\
            & \multicolumn{1}{c}{(BJD- 2455000) } & \multicolumn{1}{c}{(minutes)} & & \multicolumn{1}{c}{(BJD- 2455000) }  & 
            \multicolumn{1}{c}{(minutes)} \\
\hline
\endfirsthead

\multicolumn{3}{c}{{\tablename} \thetable{} -- Continued} \\[0.5ex]
  \hline \hline \\[-2ex]
Orbital& \multicolumn{2}{c}{Primary Eclipse} & & \multicolumn{2}{c}{Secondary Eclipse } \\
\cline{2-3} \cline{5-6}
   Cycle & \multicolumn{1}{c}{Midpoint} &  \multicolumn{1}{c}{Uncertainty } & &  \multicolumn{1}{c}{Midpoint}& \multicolumn{1}{c}{Uncertainty} \\
            & \multicolumn{1}{c}{(BJD- 2455000) } & \multicolumn{1}{c}{(minutes)} & & \multicolumn{1}{c}{(BJD- 2455000) }   & \multicolumn{1}{c}{(minutes)} \\
\hline
\endhead

 \hline
  \multicolumn{3}{l}{{Continued on Next Page\ldots}} \\
\endfoot

 \hline \hline
\endlastfoot

1	&	-32.18064 	&	0.889	& &	-24.32048	& 7.736		\\
2	&	-12.18029 	&	0.889	& &	-4.31919	&	6.602		\\
3	&	7.82051	&	0.889	& &	---	&	---	\\
4	&	27.82051	&	0.876	& &	35.67367	&	8.872		\\
5	&	47.81987	&	0.967 & &	55.68321	&	8.682		\\
6	&	67.82051	&	0.903	& &	75.66013	&	7.454		\\
7	&	87.82091	&	1.032	& &	95.68077	&	8.493		\\
8	&	107.82066	&	0.850	& &	115.67901	&	7.169		\\
9	&	127.82052	&	0.863	& &	135.67771	& 5.848		\\
10	&	147.82173	&	0.993	& &	---	&	---	\\
11	&	167.82246	&	0.850	& &	175.67458	&	6.885		\\
12	&	187.82161	&	0.876	& &	195.67849	&	6.791	\\
13	&	207.82279	&	1.229	& &	215.67642	& 7.169		\\
14	&	227.82271	&	0.889	& &	235.67813	&	6.696		\\
15	&	247.82347	& 0.889         & &	255.67295	&	7.358		\\
16	&	267.82365	&	0.876	& &	295.68469	&	7.074		\\
17	&	287.82313	& 	0.889	    & &	315.66953	& 7.839			\\
19	&	327.82365	&	0.902	    	& &	335.67685	& 6.885			\\
20	&	347.82402	&	0.889	    	& &	355.68154	& 7.764			\\
21	&	367.82395	& 0.902	    	& &	375.67816	& 7.35804			\\
22	&	387.82389	&	0.889	    	& &	395.68198	&	6.980		\\
23	&	407.82594	&	0.902	     & &	415.68345	&	6.885		\\
24	&	427.82464	&	0.876		    	& &	435.68501	&	6.980		\\
25	&	447.82520	&	0.915		    	& &	455.67885	&	6.980		\\
26	&	467.82607	&	0.889		    	& &	475.68108	& 7.169			\\
27	&	487.82642	&	0.889		    	& &	495.67777	&	7.641		\\
28	&	507.82580	&	0.941		    	& &	515.67622	&	7.930		\\
29	&	527.82476	&	1.006		    	& &	535.68698	&	7.074		\\
30	&	547.82560	&	0.889	    	& &	---	&	---	\\
31	&	---	&	---	& &	575.68315	& 	7.454	\\
32	&	587.82759	& 0.889		& &	---	&	---	\\
33	&	607.82824	&	1.006		    	& &	615.67919	&	6.696		\\
34	&	627.82612	&	0.876		    	& &	655.69609	&	7.925		\\
35	&	647.82747	&	0.876		    	& &	675.69147	&	7.650		\\
36	&	667.82816	&	0.889		    	& &	695.68003	&	7.263		\\
37	&	687.82937	&	0.889	    	& &	715.67600	&	7.454		\\
39	&	727.82918	&	0.993		    	& &	735.68862	&	7.547		\\
40	&	747.82850	&	1.006		    	& &	755.68434	&	6.980	\\
41	&	767.82945	&	0.876		    	& &	775.68358	&  7.263			\\
42	&	787.83000	&	0.967		    	& &	795.67857	&	8.211		\\
43	&	807.82964	&	0.889	& &	815.68275	&	6.980		\\
44	&	827.83067	&	0.902	& &	835.68385	&	7.265		\\
45	&	847.82987	&	0.902	& &	855.68746	&	6.696		\\
46	&	867.82980	&	0.876	& &	875.68666	&	8.114		\\
47	&	887.83118	&	0.902 & &	895.68856	&	7.547		\\
48	&	907.83190	&	0.889	& &	915.68830	&	6.696		\\
49	&	927.83150	&      0.915		& &	---	&	---	

\label{tab:eclipsetimes}
\end{longtable}

\begin{table}
\centering
\caption{Measured Radial Velocities}
\begin{tabular}{ c c c c c c c  c c }
\hline
\hline
& & &   \multicolumn{2}{c}{Broad BF} & & \multicolumn{2}{c}{Narrow BF} & \\
\cline{4-5} \cline{7-8}
Observation  & UT Date &  BJD-2455000    &  Radial &     $\sigma_{\rm{RV}}$ & & Radial &     $\sigma_{\rm{RV}}$ & $\#$ of  \\
$\#$ & &  &  Velocity  & (km s$^{-1}$) && Velocity &  (km s$^{-1}$) &  BFs \\
 & & &  (km s$^{-1}$)  & & & (km s$^{-1}$) & &   Fit \\
\hline
1& 2012 April 1 & 1019.1136730  &	34.647	&	0.111	& &	18.935	&	0.189 & 2 \\
 2 & 2012 May 26 & 1073.9616350   &	-0.050	&	0.186	& &	19.299	&	0.256 &  2 \\
 3 & 2012 May 28 & 1076.0832280    &	19.503	&	0.126	& &	19.963	&	0.218  & 2 \\
 4 & 2012 June 19 & 1098.0375080    &	32.848	&	0.176	& &	20.736	&	0.367 & 2  \\
 5 & 2012 June 19 & 1098.1085180    &	32.989	&	0.153	& &	22.249	&	0.290 & 2 \\
 6 &2012 June 20 & 1099.0462140    &	31.941	&	0.116	& &	16.303	&	0.333 & 2 \\
 7 & 2012 June 20 & 1099.0574410    &	31.925	&	0.094	& &	16.722	&	0.176 & 2 \\
 8 & 2012 July 1 & 1110.0675670     &	2.359	&	0.140	& &	20.088	&	0.133 & 2 \\
 9 & 2012 July 2 &  1111.0639150     &	-4.437	&	0.143	& &	17.350	&	0.262& 3 \\
 10 &2012 July 3 & 1112.0969640    &	-8.515	&	0.153	& &	16.860	&	0.268 & 2 \\
 11 & 2012 July 4 & 1113.0660410    &  	-5.516	&	0.275	& &	19.300	&	0.394 & 2 \\
 12 & 2012 August 4 & 1144.0612020 &	30.766	&	0.133	& &	20.612	&	0.196 & 3 \\
 
\hline
\hline
\end{tabular}
\label{tab:uncorrectedRVs}
\end{table}

\begin{table}
\centering
\caption{Preliminary ELC Radial Velocity Fit Parameters for KIC 4862625}
\begin{tabular}{ l l l  }
\hline
\hline
Parameter & Uncorrected Radial Velocities & Corrected Radial Velocities \\
\hline
K$_A$ (km s$^{-1}$) & 21.490$\pm$    0.480  & 21.185 $\pm$    0.205  \\
P (days) &  20.0002514  $\pm$ 0.0000056   & 20.0002489  $\pm$  0.0000051  \\
  T$_{conj}$ (BJD- 2455000) &  -32.18068 $\pm$    0.00010 & -32.18068 $\pm$    0.00099  \\
$e$cos($w$) &  -0.16739 $\pm$    0.00006  &  -0.16763 $\pm$      0.00010 \\
  $e$sin($w$)  &  -0.14622$\pm$     0.00093&  -0.13363$\pm$     0.00482 \\
  $e$ &  0.2223   $\pm$      0.0010  & 0.2144   $\pm$      0.0029  \\
  $w$ (degrees) & 221.14$\pm$      0.19 & 218.56  $\pm$     1.04 \\
\hline
\end{tabular}
\label{tab:rvfit}
\end{table}

\begin{table}
\centering
\caption{Corrected Radial Velocities for KIC 4862625}
\begin{tabular}{ c c c c c  }
\hline
\hline
Observation  & UT Date &  BJD-2455000    &  Radial Velocity  &     $\sigma_{\rm{RV}}$  \\
$\#$ &  & &   (km s$^{-1}$) & (km s$^{-1}$) \\
\hline
1& 2012 April 1 & 1019.1136730   &   34.746   &   0.111 \\
 2 & 2012 May 26 & 1073.9616350   &   -0.314     &  0.186  \\
 3 & 2012 May 28 & 1076.0832280    &  18.575   &   0.126 \\
 4 & 2012 June 19 & 1098.0375080    &  31.146    &  0.176   \\
 5 & 2012 June 19 & 1098.1085180    &  29.775    &  0.153 \\
 6 &2012 June 20 & 1099.0462140    &  34.672    &  0.116  \\
 7 & 2012 June 20 & 1099.0574410    &  34.237    &  0.094 \\
 8 & 2012 July 1 & 1110.0675670     &  1.306    &  0.140  \\
 9 & 2012 July 2 &  1111.0639150     & -2.752    &  0.143  \\
 10 &2012 July 3 & 1112.0969640     & -6.340    &  0.153  \\
 11 & 2012 July 4 & 1113.0660410    &  -5.782   &   0.275  \\
 12 & 2012 August 4 & 1144.0612020    &  29.189    &  0.132  \\
 
\hline
\hline
\end{tabular}
\label{tab:RVs}
\end{table}

\begin{landscape}

\begin{longtable}{rllll}

 \caption[short]{ Model parameters for PH1b. The reference epoch is $t_0 = $2,454,970 (BJD). $\dagger$ - parameter median and credible intervals determined from a combination of EB-only and EB-fixed, planetary solutions.   \label{tab:Parameters} } \\ 
\hline
\hline
Index & Parameter Name & Joint Solution & EB-only & Planet only (EB-Fixed) \\
\endfirsthead

\hline
\hline
Index & Parameter Name & Joint Solution & EB-only & Planet only (EB-Fixed) \\
\hline
\endhead

 \hline
  \multicolumn{3}{l}{{Continued on Next Page\ldots}} \\
\endfoot

 \hline \hline
\endlastfoot

\hline
& ~{\bf Mass parameters} & & &  \\
0 &~~Mass of Star Aa, $M_{Aa}$ ($M_\odot$) & $                                    1.384\pm                                    0.079$ & -- & $                                    1.528\pm                                    0.087$$\dagger$\\
1 &~~Mass Ratio, $M_{Ab}/M_{Aa}$ & $                                   0.2794\pm                                   0.0051$ & -- & $                                   0.2663_{-                                  0.0023}^{+                                  0.0033}$\\
2 &~~Planet Mass Ratio, $M_b/M_{Aa}$ (99.7\% conf.) & $ <                                 0.000357$ & $ 0$ (fixed) & $ 0$ (fixed)\\
&~{\bf Planetary Orbit} & & & \\
3 &~~Period, $P_b$ (days) & $                                  138.506_{-                                   0.092}^{+                                   0.107}$ & -- & $                                  138.317_{-                                   0.027}^{+                                   0.040}$\\
4 &~~Eccentricity Parameter, $\sqrt{e_b} \cos(\omega_b)$ & $                                    0.228_{-                                   0.014}^{+                                   0.012}$ & -- & $                                   0.2519_{-                                  0.0060}^{+                                  0.0044}$\\
5&~~Eccentricity Parameter, $\sqrt{e_b} \sin(\omega_b)$ & $                                   -0.044_{-                                   0.023}^{+                                   0.034}$ & -- & $                                  -0.0824_{-                                  0.0044}^{+                                  0.0061}$\\
6&~~Bary. Transit Time, $t_b$ (BJD-2,454,900) & $                                   174.71\pm                                     0.11$ & -- & $                                  174.490_{-                                   0.036}^{+                                   0.052}$\\
7&~~Sky-plane Inclination, $i_b$ (deg) & $                                   90.022_{-                                   0.061}^{+                                   0.072}$ & -- & $                                   90.050\pm                                    0.053$\\
8 &~~Relative Nodal Angle, $\Delta\Omega$ (deg) & $                                     0.89\pm                                     0.16$ & -- & $                                     0.89\pm                                     0.14$\\
&~{\bf Stellar Orbit} & & & \\
9 &~~Period, $P_A$ (days) & $                                20.000214_{-                                0.000043}^{+                                0.000025}$ & $                               20.0002468\pm                                0.0000044$ & $          20.00024678$ (fixed)\\
10 &~~Eccentricity Parameter, $e_A \cos(\omega_A)$ & $                                 -0.16773_{-                                 0.00018}^{+                                 0.00021}$ & $                                 -0.16747\pm                                  0.00037$ & $             -0.16750$ (fixed)\\
11 &~~Eccentricity Parameter, $e_A \sin(\omega_A)$ & $                                  -0.1291\pm                                   0.0084$ & $                                   -0.141\pm                                    0.015$ & $              -0.1393$ (fixed)\\
12 &~~Primary Eclipse Time, $t_A$ (BJD-2,454,900) & $                                 67.81776\pm                                  0.00015$ & $                                 67.81797\pm                                  0.00016$ & $             67.81798$ (fixed)\\
13&~~Sky-plane inclination, $i_A$ (deg) & $                                   87.360_{-                                   0.072}^{+                                   0.063}$ & $                                   87.592\pm                                    0.053$ & $                87.58$ (fixed)\\
&~{\bf Radius/Light Parameters} & & &  \\
14&~~Linear Limb Darkening Parameter, $u_{Aa}$ & $                                     0.27\pm                                     0.15$ & $                                     0.31_{-                                    0.21}^{+                                    0.29}$ & $ 0.32$ (fixed)\\
15&~~Quadratic Limb Darkening Parameter, $v_{Aa}$ & $                                     0.08\pm                                     0.18$ & $                                     0.17_{-                                    0.28}^{+                                    0.21}$ & $ 0.17$ (fixed)\\
16&~~Stellar Flux Ratio, $F_{Ab}/F_{Aa}$ $\times 100$)& $                                   0.1533_{-                                  0.0099}^{+                                  0.0142}$ & $                                   0.1414_{-                                  0.0036}^{+                                  0.0040}$ & $ 0.1405$ (fixed)\\
17 &~~Density of Star Aa, $\rho_{Aa}$ (g/cm$^3$) & $                                   0.2542\pm                                   0.0076$ & -- & $                                    0.293\pm                                    0.013$$\dagger$\\
18 &~~Radius Ratio, $R_{Ab}/R_{Aa}$ & $                                    0.240_{-                                   0.015}^{+                                   0.020}$ & $                                    0.219\pm                                    0.014$ & $ 0.215$ (fixed)\\
19& ~~Planet Radius Ratio, $R_b/R_{Aa}$ & $                                  0.03222\pm                                  0.00045$ & -- & $                                  0.03268\pm                                  0.00032$\\
20 &~~Contamination, $F_X/F_{Aa}$ ($\times 100$) & $                                      7.5\pm                                      1.9$ & $                                     11.8\pm                                      1.9$ & $   12.$ (fixed)\\
&~{\bf Noise Parameter} & & &  \\
21&~~Long Cadence Relative Width, $\sigma_{LC}$ ($\times 10$$^5$ ) &$ 20.33 \pm                                      0.20$ &   $20.31_{-                                 0.21}^{+                                    0.20}$ & 20.31(fixed)\\                               \\

&~{\bf Radial Velocity Parameters} & &  & \\
22&~~RV Offset $\gamma$ (km/s) & $                                     18.18_{-                                    0.24}^{+                                    0.25}$ & $                                     18.09_{-                                 0.27}^{+                                    0.29}$ & 18.09 (fixed) \\
23&~~RV Jitter (km/s) & $                                     0.80_{-                                    0.17}^{+                                    0.28}$ & $                                     0.81_{-                                    0.18}^{+                                    0.27}$ & --\\

&~~Mass Function$^*$, $f(M_{Aa},M_{Ab})$ ($M_\odot$) & $                                  0.01836\pm                                  0.00082$ & $                                  0.01802\pm                                  0.00089$ & $              0.01807$ (fixed)\\
&~~RV Semi-amplitude$^*$, $K_{Aa}$ (km/s) & $                                    20.69\pm                                     0.31$ & $                                    20.56\pm                                     0.33$ & --\\

\hline
\hline
\caption{ Model parameters for PH1 system. The reference epoch is $t_0 = $2,454,970 (BJD). $\dagger$ - parameter median and credible intervals determined from a combination of EB-only and EB-fixed, planetary solutions.  $^*$ - parameter fit for in the EB-only solution and derived from other fit parameters in the joint solution. Reported uncertainties  represent the 68$\%$ confidence intervals and correspond to $\pm$ 1-$\sigma$ when the marginalized posterior is Gaussian. \label{tab:Parameters}}
\end{longtable}
\end{landscape}

\begin{table}
\caption{Derived planet and binary parameters from photometric-dynamical model fit. Reference epoch is t$_{0}$ =2454970 (BJD) .}
\begin{tabular}{lllll}
\hline
\hline
Parameter Name & Joint Solution & EB-only & Planet only (EB-Fixed) \\ \hline
{\bf Stellar (Aa $\&$ Ab) Bulk parameters} & & & \\
Normalized Radius, $R_{Aa}/a_A$ & $                                  0.04690\pm                                  0.00041$ & $                                 0.04490\pm                                  0.00064$ & $              0.04516$ (fixed)\\
Radius of Star Ab, $R_{Ab}$ ($R_\odot$) & $                                    0.422_{-                                   0.029}^{+                                   0.038}$ & -- & $                                    0.378\pm                                    0.023$$\dagger$\\
Mass of Star Ab, $M_{Ab}$ ($M_\odot$) & $                                    0.386\pm                                    0.018$ & -- & $                                    0.408\pm                                    0.024$$\dagger$\\
Density of Star Ab, $\rho_{Ab}$ (g/cm$^3$) & $                                      5.1\pm                                      1.2$ & -- & $                                      7.5_{-                                     1.3}^{+                                     1.4}$$\dagger$\\
Gravity of Star Aa, $\log(g_{Aa})$ (cgs) & $                                    4.089\pm                                    0.014$ & -- & $                                    4.144\pm                                    0.014$$\dagger$\\
Gravity of Star Ab, $\log(g_{Ab})$ (cgs) & $                                    4.772_{-                                   0.071}^{+                                   0.061}$ & -- & $                                    4.892\pm                                    0.057$$\dagger$\\
{\bf Stellar (Aa $\&$ Ab) Orbital Properties} & & & \\
Semimajor axis , $a_{A}$ (AU) & $                                   0.1744\pm                                   0.0031$ & -- & $                                   0.1797\pm                                   0.0035$$\dagger$\\
Eccentricity, $e_A$ & $                                   0.2117\pm                                   0.0051$ & $                                   0.2188\pm                                   0.0094$ & $               0.2179$ (fixed)\\
Argument of Periastron $\omega_A$ (deg) & $                                    217.6\pm                                      1.9$ & $                                    220.1_{-                                     3.2}^{+                                     2.9}$ & $             219.7504$ (fixed)\\
{\bf Planet Bulk Properties} & & & \\
Planet Radius, $R_b$ ($R_\oplus$) & $                                     6.18\pm                                     0.14$ & -- & $                                    6.18\pm                                     0.17$$\dagger$\\
Planet Mass, $M_b$ ($M_\oplus$, 99.7\% conf.) & $ <                                     169.$ & -- & --\\

{\bf Planet Orbital Properties} & & & \\
Semimajor axis $a_{b}$ (AU) & $                                    0.634\pm                                    0.011$ & -- & $                                    0.652\pm                                    0.012$$\dagger$\\
Eccentricity of Planetary Orbit, $e_b$ & $                                   0.0539\pm                                   0.0081$ & -- & $                                   0.0702_{-                                  0.0039}^{+                                  0.0029}$\\

Argument of Periastron, $\omega_b$ (deg) & $                                    348.0_{-                                     5.0}^{+                                     6.7}$ & -- & $                                   341.88_{-                                    0.70}^{+                                    0.97}$\\
Mutual Inclination$^a$, $I$(deg) & $                                    2.814\pm                                    0.073$ & -- & $                                    2.619\pm                                    0.057$\\

\hline
\hline
\end{tabular}
\tablenotetext{a}{Stellar to planetary orbit mutual inclination: The mutual inclination is the angle between
the orbital planes of the binary and the planet, and is
defined as
$\cos I  =  \sin i_A \sin i_b \cos \Delta \Omega 
+ \cos i_A \cos i_b$.  $\dagger$ - parameter median and credible intervals determined from a combination of EB-only and EB-fixed, planetary solutions. Reported uncertainties represent the 68$\%$ confidence intervals and correspond to $\pm$ 1-$\sigma$ when the marginalized posterior is Gaussian.  }

\label{tab:DerivedParameters}
\end{table}

\begin{deluxetable}{ll}
\tablecaption{Parameters from full ELC Model}
\tablewidth{0pt}
\tablehead{
\colhead{Parameter} &          
\colhead{Best fit}
}
\startdata
$K_{A}$ (km s$^{-1}$) & $   21.179 \pm 0.162$ \\
  $e$                 & $0.2165 \pm 0.0013$ \\
$\omega$ (deg)      & $219.30\pm 0.45$ \\
$R_{Aa}/a$             & $ 0.04489 \pm 0.00007$\\
$R_{Ab}/a$             & $ 0.00938 \pm       0.00008$ \\
$T_{\rm eff, Ab}/T_{\rm eff, Aa}$           & $0.5557\pm         0.0062$  \\
$i$ (deg)           & $87.612  \pm    0.004$ \\
$u_Aa$               & $0.208 \pm       0.020$ \\
$v_Aa$               & $0.293\pm       0.0249$  \\
$P$ (days)          & $20.0002490 \pm 0.0000049$ \\
  $T_{\rm conj}$ (BJD-2455000)   &  -32.18064 $\pm$ 0.00203  \\
Contamination & 0.109 $\pm$   0.012       \\
\enddata
\label{tab:ELC}
\tablecomments{
Note: Subscript ``Aa'' denotes the primary star, subscript ``Ab'' the
secondary star in the eclipsing binary KIC4862625 .}
\end{deluxetable}

\clearpage

\clearpage

\end{document}